\pdfoutput=1
\documentclass[american,aps,pra,twocolumn,reprint,superscriptaddress,floatfix,10pt]{revtex4-2}
\usepackage[utf8]{inputenc}
\usepackage{color}
\usepackage[american]{babel}
\usepackage{units}
\usepackage{amsmath}
\usepackage{xcolor}
\usepackage{amsthm}
\usepackage{amssymb}
\usepackage{graphicx}
\usepackage{comment}
\usepackage[unicode=true,pdfusetitle,
 bookmarks=false,bookmarksnumbered=false,bookmarksopen=false,
 breaklinks=true,pdfborder={0 0 0},pdfborderstyle={},backref=false,colorlinks=true]
 {hyperref}
\hypersetup{colorlinks,citecolor=blue,linkcolor=blue,urlcolor=blue,hypertexnames=true}



\global\long\def\ket#1{\left|#1\right\rangle }%
\global\long\def\bra#1{\left\langle #1\right|}%
\global\long\def\expected#1{\left\langle #1\right\rangle }%

\begin{abstract}
We obtain the ground-state phase diagram of two spin chains consisting in a set two-level systems asymmetrically coupled to an XX chain through a chiral interaction. The interaction is parametrized by its magnitude and an angle defined by the relative orientation of the spins in different chains. From the entanglement spectrum, we identify the critical lines separating distinct magnetically ordered phases, with the interaction angle able to shift or fully suppress the transition. By increasing the coupling strength, the systems is driven through a quantum phase transition, leading to the formation of two types of in-plane antiferromagnetic stripes. The interaction strength sets stripe formation, while the angle controls the spins orientations. The chiral interaction also induces a non-trivial finite vector spin chirality with opposite orientation on the chains. We show that the vector spin chirality emerges smoothly from the decoupled limit and occurs for angles different from zero and $\pi/2$, where collinear order is favored instead.
\end{abstract}

\begin{document}

\title{Tunable Magnetic Order in Chiral Coupled Spin Chains}

\author{Rafael D. Soares}
\affiliation{Max Planck Institute for the Physics of Complex Systems, N\"{o}thnitzer Stra{\ss}e 38, 01187 Dresden, Germany}
\affiliation{Departamento de Física e Astronomia, Faculdade de Ciências da Universidade do Porto, Rua do Campo Alegre, s/n, 4169-007 Porto, Portugal}

\author{J. M. Viana Parente Lopes}
\affiliation{Departamento de Física e Astronomia, Faculdade de Ciências da Universidade do Porto, Rua do Campo Alegre, s/n, 4169-007 Porto, Portugal}
\affiliation{Centro de Física das Universidades do Minho e do Porto (CF-UM-UP) and Laboratório de Física para Materiais e Tecnologias Emergentes LaPMET, Universidade do Porto, 4169-007 Porto, Portugal}

\author{Hugo Terças}
\affiliation{Instituto Superior de Engenharia de Lisboa, Instituto Politécnico de Lisboa, Portugal}
\affiliation{GoLP/Instituto de Plasmas e Fusão Nuclear, Instituto Superior Técnico, Universidade de Lisboa, Portugal}

\date{\today}
\maketitle

\section{Introduction}
Quantum spin chains are a cornerstone of quantum many-body physics. Despite their apparent simplicity, they host magnetically ordered states, topological phases with fractionalized excitations~\cite{PhysRevLett.50.1153,PhysRevLett.65.3181,Mishra2021}, and quantum paramagnets with long-range entanglement such as spin liquids~\cite{PhysRevB.65.165113,RevModPhys.89.041004,doi:10.1126/science.aay0668}. Their richness has driven major theoretical advances, in part because many are exactly solvable in one dimension. Classic examples include the transverse-field Ising model (TFIM), which can be mapped to free fermions via a Jordan-Wigner (JW) transformation~\cite{Sachdev2011}, and the XXZ chain, solved by the Bethe ansatz~\cite{Bethe1931,Yang1969}. Such models have shaped our understanding of correlation functions, conformal field theory, and integrability~\cite{Itzykson1989,Fradkin2013}. Other one-dimensional spin-chain models have illuminated more exotic forms of quantum order: the Affleck--Kennedy--Lieb--Tasaki (AKLT) chain realizes a gapped Haldane phase with spin-$\nicefrac{1}{2}$ edge states, while the Majumdar--Ghosh chain hosts a dimerized ground state at a special coupling ratio.  

Beyond exactly solvable models, progress has been driven by state-of-the-art numerical algorithms and, in parallel, by quantum simulation platforms. Specifically, advances in tensor-network methods~\cite{Ors2014}, grounded in the area-law scaling of entanglement in low-dimensional systems~\cite{Hastings2007,PhysRevLett.100.070502}, have established matrix product states (MPS) as efficient variational \emph{ansatz} for one-dimensional ground states~\cite{PhysRevLett.75.3537,PhysRevLett.100.030504}. The Density Matrix Renormalization Group (DMRG)~\cite{White1992,PhysRevB.48.10345,RevModPhys.77.259}, formulated as a variational optimization over MPS~\cite{Schollwck2011}, remains the method of choice for one-dimensional spin chains, capturing both gapped and critical phases. Complementing these theoretical tools, highly controlled experimental platforms, particularly those based on ultracold atoms in optical lattices, have enabled the realization of spin-$1/2$ chains with tunable interactions~\cite{Grg2019}. These setups have allowed the direct observation of quantum criticality and phase transitions~\cite{Bloch2008,Islam2015}, while quantum gas microscopy provides real-space imaging of spin correlations and dynamics~\cite{Parsons2016,Cheuk2016}, offering direct probes of spin~\cite{Parsons2016,Cheuk2016} and charge-ordered~\cite{Hirthe2023,https://doi.org/10.48550/arxiv.2412.17801} phases.  

In recent years, considerable effort has been devoted in understanding the role of chiral interactions in quantum spin chains. Of particular relevance to this work are the chiral spin networks~\cite{Ramos2014,Pichler2015,MartnezPea2020,Kora2024}, which consist of ensembles of localized two-level atoms coupled to a chiral environment. Such environments break inversion and time-reversal symmetry, giving rise to distinct propagation velocities for excitations moving in opposite directions. Beyond their fundamental interest, these platforms enable a wide range of applications, from quantum information processing architectures~\cite{Mahmoodian2016,Wang2022,Palaiodimopoulos2024} to dissipative state preparation~\cite{Stannigel2012,Ramos2014,Barik2018,Hei2024} and the exploration of non-reciprocal many-body phenomena~\cite{Nadolny2025,PhysRevLett.132.120401,soares2025dissipativephasetransitioninteracting}. Various physical implementations have been proposed, including optical fibers~\cite{PhysRevLett.110.243603,PhysRevLett.113.143601}, photonic waveguides, and magnonic devices~\cite{PhysRevA.93.062104,Hei2024}. The latter can be effectively modeled by a non-interacting spin chain coupled to an XX chain by direction-dependent complex couplings, where neighboring interactions differ by a phase factor. Additional experimental platforms include ultracold atoms with synthetic gauge fields~\cite{Goldman2014,Goldman2016,Lienhard2020,Eix2025}, Rydberg atom arrays~\cite{Kuznetsova2023}, and quantum optical architectures~\cite{Zoller2012,Vermersch2016,Gross2017}, all of which offer the tunability and coherence required to test theoretical predictions.

Most previous works have focused on the dynamics of excitations in non-interacting spin systems and on the emergence of novel non-equilibrium quantum phases~\cite{Stannigel2012,Ramos2014} in their steady states, where the time evolution is typically described by master equations obtained after tracing out the photonic or spin-waveguide environment. In contrast, here we study in detail the ground-state phase diagram of the full system, highlighting the non-trivial role of the complex phase factor in shaping the underlying order. Specifically, we consider two spin-$1/2$ chains arranged in a triangular-lattice geometry and coupled via a complex interchain interaction. This interaction is characterized by its strength $g$ and a geometric angle $\phi$ (the phase of the complex coupling), with a sign that depends on the position of the nearest-neighbor spin, as shown in Fig.~\ref{full_scheme}. This setup mimics a minimal chiral spin network: the first chain consists of independent spins, while the second is modeled by an XX chain, allowing for the propagation of spin excitations. By mapping the model to fermions, we analyze the entanglement spectrum and its topological properties. We show, both numerically and analytically, that not only the coupling strength but also the geometric angle shifts the position of the critical lines and, in some cases, even completely suppresses the quantum phase transition. By extracting the relevant critical exponents, we establish that the system’s critical behavior falls within the Ising universality class, regardless of the geometric phase. Finally, by inspecting the spin–spin correlation function, we demonstrate that the model realizes a rich phase diagram featuring regions with collinear order, specifically, antiferromagnetic in-plane stripe order whose precise orientation can be tuned through the interaction angle, as well as, non-collinear orders in which each chain exhibits a finite vector spin chirality.

The paper is organized as follows: in Section~\ref{sec:Hamiltonian_model}, we introduce the Hamiltonian of the model and the specific form of the chiral interaction. Section~\ref{sec:boundaries_phase_diagram} presents the mapping of the spin model to fermions, from which we derive, under suitable approximations, an analytic expression for the critical lines of the phase diagram, complemented by a numerical analysis of the entanglement-spectrum degeneracies. In Section~\ref{sec:magnetic_order_universality_class}, we investigate the in-plane correlation functions and extract the universal critical exponents of the quantum phase transition. We then examine the distinct types of collinear and non-collinear magnetic order across different parameter regimes, with particular emphasis on the role of the geometry angle as a tunable parameter controlling the chiral interactions and mediating between competing magnetic orders. Finally, Section~\ref{sec:conclusions} summarizes our findings. Several appendices provide additional methodological details and supporting results.

\section{Model and Methods}
\label{sec:Hamiltonian_model}
\begin{figure}[t]
\begin{centering}
\includegraphics{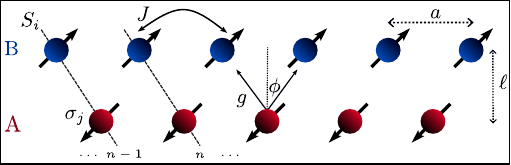}
\par\end{centering}
\caption{(color online) Schematic representation of the coupled spin chains. Both operators $\sigma$ (chain A) and $S$ (chain B) represent spin-$\nicefrac{1}{2}$ operators with an energy level splitting $\omega_0$ and $\Omega_0$, respectively. In chain B, the excitations can hop at a rate of $J$. The chains are coupled through a phase-dependent $\left(\phi\right)$ interaction with a strength $g$. In both chains, the spins are equally spaced at a distance $a$.}
\label{full_scheme}
\end{figure}
We consider two interlinked quantum spin-\nicefrac{1}{2} chains: chain A $\left(\sigma_{j}\right)$ is a set of $N$ independent two-level systems, while chain B $\left(S_{i}\right)$ is a set of $N$ two-level systems with a nearest-neighbor XY exchange interaction. The two chains are coupled such that their geometric arrangement forms a triangular-shaped spin ladder. Each spin in chain A ($\sigma_{n}$) is assumed to interact with its nearest neighbors, specifically $S_n$ and $S_{n+1}$. The coupling between the two chains is designed so that the interaction depends on the geometric angle $\phi$, defined by the relative positions of the spins, as illustrated in Fig.~\ref{full_scheme}. Consequently, $\phi$ is restricted to lie between 0 and $\pi/2$. The system is described by the Hamiltonian of each chain plus a Hamiltonian characterizing their coupling: $\mathcal{H}=\mathcal{H}_{\text{A}}+\mathcal{H}_{\text{B}}+\mathcal{H}_{\text{int}}$. 
The Hamiltonian of the spin chain A corresponds to
\begin{equation}
\mathcal{H}_{\text{A}}=\omega_{0}\sum_{n=0}^{N-1} \sigma_{n}^{+}\sigma_{n}^{-},\label{eq:system_hamiltonian}
\end{equation}
where $\sigma^{+(-)}_n$ corresponds to a rising (lowering) operator spin-\nicefrac{1}{2} operator on site $n$ and $\omega_{0}$ is the energy level splitting. While for chain B we have
\begin{equation}
\mathcal{H}_{\text{B}}=J\sum_{n=0}^{N-2}\left( S_{n}^{+}S_{n+1}^{-}+ S_{n}^{-}S_{n+1}^{+} \right)+\Omega_{0}\sum_{n=0}^{N-1} S_{n}^{+}S_{n}^{-},\label{eq:hamiltonian_reservoir}
\end{equation}
with $\Omega_{0}$ the energy level splitting and $J$ the strength of the nearest-neighbor XY exchange interaction.

As described above, we consider an interaction Hamiltonian that depends on the geometric arrangement of the chains, encoded through a phase-dependent coupling of the form $g(\phi) = g e^{i\phi}$. The interaction reads  
\begin{equation}
\mathcal{H}_{\text{int}} = 2g \sum_{n=0}^{N-2} \sigma_n^+ \left( e^{i\phi} S_{n+1}^x + e^{-i\phi} S_n^x \right) + \text{h.c.},  
\label{eq:interaction_hamiltonian}
\end{equation}
where $g$ is the coupling strength and $S_n^x$ denotes the spin-$1/2$ operator along the $x$-axis at site $n$. For generality, we retain the counterrotating terms $(\sigma^+ S^+)$, making the model applicable across different energy hierarchies. As a result, the continuous spin U(1) symmetry associated with the conservation of the total number of excitations, $\mathcal{N} = \sum_n \sigma_n^+ \sigma_n^- + \sum_n S_n^+ S_n^-,$ is explicitly reduced to a discrete $\mathbb{Z}_2$ symmetry. The Hamiltonian is no longer invariant under arbitrary global spin rotations about the $z$-axis, but remains invariant under a rotation by $\pi$: $S^\pm \; (\sigma^\pm) \rightarrow -S^\pm \; ( -\sigma^\pm),$ generated by the spin parity operator  
\begin{equation}
\mathcal{P} = \exp\left( i\pi \sum_{n=0}^{N-1} \left[ \sigma_n^+ \sigma_n^- + S_n^+ S_n^- \right] \right).  
\label{eq:symmetry_op}
\end{equation}
Thus, while the total excitation number is not conserved, its parity remains a good quantum number.

More importantly, the interaction is chiral, as it explicitly violates both time-reversal and inversion symmetry, except at $\phi = 0$, where time-reversal symmetry is restored. Moreover at \(\phi = \pi/2\), the system is symmetric under the combination of time-reversal with spin parity applied only to chain B.

To study the zero-temperature phase diagram of the system, we first do an analytical treatment by (under a certain approximation) mapping it into a chain of spinless fermions using the JW transformation. To further investigate the system's ground-state phase diagram, we computed the spin-spin correlation functions, low-lying energy levels, the entanglement entropy and the entanglement spectra by doing DMRG~\cite{White1992, Schollwoeck2011} calculations as implemented in the ITensor Library~\cite{Fishman2022,Fishman2022a}. We performed our calculations on systems with up to $4096$ spins under open-boundary conditions (OBC). Throughout the simulations, we maintained a truncation error below $10^{-10}$ and stopped the sweeping process only once the specific convergence criteria were met. In particular, when computing the energy gap, we ensured that the energy variance of the obtained eigenstates remained below $10^{-6}$. Our study concentrates exclusively on the regime where both $\Omega_{0},\omega_{0}>2J$. This ensures that in the non-interacting limit ($g=0$), the system has gapped excitations and a unique ground state: the completely polarized down-state $\ket{\text{GS}}=\ket{\downarrow\cdots\downarrow}$.

\section{Boundaries of the Phase Diagram}
\label{sec:boundaries_phase_diagram}
To identify the critical lines that mark the phase boundaries in our model and to analyze their dependence on the geometric interaction angle $\phi$, we first map the spin Hamiltonian to a fermionic one using the JW transformation~\cite{Jordan1928,Greiter2014}. This transformation establishes a correspondence between the spin operators on the two chains and fermionic creation and annihilation operators, which we construct as follows:
\begin{equation}
\begin{aligned}
\sigma_{n}^{+} & =c_{n}^{\dagger}\exp\left(-i\pi\left[\sum_{l=0}^{n-1}\left[b_{l}^{\dagger}b_{l}+c_{l}^{\dagger}c_{l}\right]+b_{n}^{\dagger}b_{n}\right]\right),\\
\sigma_{n}^{z} & =c_{n}^{\dagger}c_{n}-\nicefrac{1}{2},\\
S_{n}^{+} & =b_{n}^{\dagger}\exp\left(-i\pi\sum_{l=0}^{n-1}\left[b_{l}^{\dagger}b_{l}+c_{l}^{\dagger}c_{l}\right]\right),\\
S_{n}^{z} & =b_{n}^{\dagger}b_{n}-\nicefrac{1}{2},
\end{aligned}
\end{equation}
where $c_{n}^{\dagger}\left(b_{n}^{\dagger}\right)$ and $c_{n}\left(b_{n}\right)$ are the fermionic operators that create/destroy a fermion in the nth site of chain A(B). In this transformation, the interaction terms $\sigma^+ S^+$ correspond to the pairing terms $b^\dagger c^\dagger$, while the XY exchange couplings $\sigma^+ S^-$  lead to hoppings $b^\dagger c$. This mapping avoids the typical issues of spin-wave approaches, where the counterrotating interaction leads to bosonic pairing terms which may originate dynamical instabilities~\cite{Wang2019}, as we explore and compare in appendix~\ref{sec:SPIN-WAVE-THEORY}. The fermionic Hamiltonian is then given by
\begin{equation}
\begin{aligned}\mathcal{H}_{\text{A}} & =\omega_{0}\sum_{n=0}^{N-2}c_{n}^{\dagger}c_{n},\\
\mathcal{H}_{\text{B}} & = \sum_{n=0}^{N-1}\Omega_{0}b_{n}^{\dagger}b_{n}+J\sum_{n=0}^{N-2}\left(b_{n+1}^{\dagger}b_{n}+\text{h.c}\right)e^{i\pi c^\dagger_n c_n},\\
\mathcal{H}_{\text{int}} & =g\sum_{n=0}^{N-2}\left[e^{i\phi}c_{n}^{\dagger}\left(b_{n+1}^{\dagger}+b_{n+1}\right)+\right.\\
 & \qquad\qquad\qquad+\left.e^{-i\phi}c_{n}^{\dagger}\left(b_{n}-b_{n}^{\dagger}\right)+\text{h.c.}\right],
\end{aligned}
\label{eq:fermionic_hamiltonians}
\end{equation}
where $\exp (i\pi c^\dagger_n c_n)=1-2c^\dagger_n c_n$, so the Hamiltonian contains an interaction component that reads as
\begin{equation}
   \mathcal{H}^\text{quartic}_B= -2J\sum_{n=0}^{N-2}\left(b_{n+1}^{\dagger}b_{n}+\text{h.c}\right)c^\dagger_n c_n.
   \label{eq:quartic_terms}
\end{equation}
The spin Hamiltonian is only mapped to a quadratic fermionic Hamiltonian when the two chains are decoupled ($g=0$)\footnote{In this case, one can make the JW separately on each chain.} or when the hopping in chain B, $J$, is considerably smaller than other energy scales, i.e, $J \ll g,\Omega_{0},\omega_{0}$.

\subsection{Free-Fermionic Approximation}
To make analytical progress, we first neglect the quartic term in $\mathcal{H}_B$ of Eq.~\eqref{eq:fermionic_hamiltonians}, an approximation that we expect to be valid when $g>J$. Assuming periodic boundary conditions (PBC) and introducing the Nambu spinor $\Psi^{\dagger}_k = \begin{pmatrix}b_{k}^{\dagger} & c_{k}^{\dagger} & b_{-k} & c_{-k}\end{pmatrix}$, we can express the quadratic part of the fermionic Hamiltonian in Bogoliubov-de-Gennes (BdG) form~\cite{Sato2017},
\begin{equation}
\mathcal{H}=\dfrac{1}{2} \sum_{k \in \mathcal{K}_p} \Psi_{k}^{\dagger}H_k\Psi_{k}+E_{0},\label{eq:Bulk_fermion_bogo_de_gennes}
\end{equation}
where $\mathcal{K}_p$ denotes the set of allowed quasi-momenta $k$, $E_{0}=J\sum_{k \in \mathcal{K}_p}\cos\left(k\right)+N\left(\omega_{0}+\Omega_{0}\right)/2$ and
\begin{equation}
H_k=\begin{pmatrix}\Omega_{k } & \Upsilon_{k}^{\ast} & 0 & -\gamma_{-k}\\
\Upsilon_{k} & \omega_{0} & \gamma_{k} & 0\\
0 & \gamma_{k}^{\ast} & -\Omega_{k} & -\Upsilon_{-k}\\
-\gamma_{-k}^{\ast} & 0 & -\Upsilon_{-k}^{\ast} & -\omega_{0}
\end{pmatrix},\label{eq:single_particle_bdg}
\end{equation}
with 
\begin{equation*}
\begin{aligned}
\Omega_k&= \Omega_{0} + 2J\cos(k), \\
\gamma_{k}&=2ige^{-ik/2}\sin\left(\phi-\frac{k}{2}\right),\\
\Upsilon_{k}&=2ge^{-ik/2}\cos\left(\phi-\frac{k}{2}\right).
\end{aligned}
\end{equation*} 
When applying the JW transformation with PBC in the spin model, special care must be taken with the boundary terms. In particular, the JW string, $(\exp (-i \pi \sum_l c^\dagger_l c_l + b^\dagger_l b_l))$, generates an additional contribution in the transformation of the boundary terms. However, since we are interested in the physics in the thermodynamic limit, we neglect this difference and consistently use PBC in the fermionic model, as the boundary term introduces only corrections of order $\mathcal{O}\left(1/N \right)$.

\begin{figure}[t]
\centering
\includegraphics{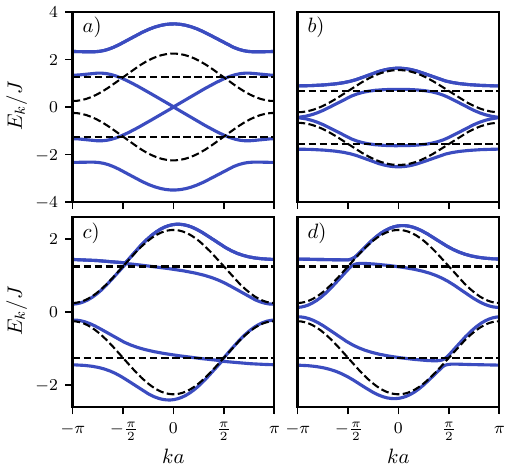}
\caption{(color online) Single-particle band structure of the Hamiltonian in Eq.~\ref{eq:single_particle_bdg} for different strength and angle of interaction (solid blue lines). The dashed black curves represents the bare dispersion (when $g=0$). In panel $a)$, the parameters are $\left(g,\phi\right)=(1.677 J,0)$; in panel $b)$, $\left(g,\phi\right)=(0.559 J, \pi/2)$; in panel $c)$, $\left(g,\phi\right)=\left(0.5 J, \pi/4 \right)$; and in panel $d)$, $\left(g,\phi\right)=\left(0.5, \pi/3 \right)$. Other parameters: $\Omega_{0}=\omega_{0}=2.5J$.\label{fig:Hybridisation-fermions}}
\end{figure}

In Fig.~\ref{fig:Hybridisation-fermions}, we show the single-particle dispersion relation obtained by diagonalizing the matrix in Eq.~\ref{eq:single_particle_bdg}. The system supports both gapless (see panels $a)$ and $b)$) and gapped phases (check panels $c)$ and $d)$). Every positive energy eigenstate in the spectrum has a corresponding partner with symmetric energy and momentum, which results from the single-particle Hamiltonian in Eq~\eqref{eq:single_particle_bdg} satisfying the particle-hole constraint~\cite{Sato2017,RevModPhys.88.035005} (PHC) typical of BdG Hamiltonians,
\begin{equation}
    \mathcal{X} \mathcal{H}_k \mathcal{X}^{-1} = - \mathcal{H}_{-k},
    \label{eq:phc}
\end{equation}
where $\mathcal{X}=\left(\sigma^x \otimes \mathcal{I}_{2\times 2}\right) \mathcal{K}$ with $\sigma^x$ the $x$ Pauli matrix, $\mathcal{I}_{2\times 2}$ the $2\times 2$ identity matrix and $\mathcal{K}$ complex conjugation operator. The angle $\phi$ is analogous to a Peierls phase that arises when a magnetic field is incorporated into a tight-binding model. So, the time-reversal symmetry is explicitly broken for $\phi \neq 0$. As inversion symmetry is also explicitly broken by the interaction, this results in an asymmetry in the bands, as seen in panels $c)$ and $d)$ of Fig.~\ref{fig:Hybridisation-fermions} (for a given band index $\lambda$: $E_{\lambda,k} \neq E_{\lambda,-k}$). For $\phi = \pi/2$, the Hamiltonian has an additional anti-unitary symmetry, corresponding to the combination of time-reversal symmetry with the fermionic parity transformation acting on only one of the chains. At the level of the single-particle Hamiltonian, this is implemented by the matrix $\mathcal{S} = \left(\mathcal{I}_{2\times 2} \otimes \sigma^z \right) \mathcal{K}$  where $\sigma^z$ is the $z$-Pauli matrix, and it imposes the constraint
\begin{equation}
\mathcal{S} H_k \mathcal{S}^{-1} = H_{-k}.
\label{eq:anti-unitary_symmetry}
\end{equation}
This implies that for $\phi = \pi/2$, every energy eigenstate in the spectrum has a corresponding partner with opposite momentum, which is indeed observed in panel $b)$ of Fig.~\ref{fig:Hybridisation-fermions}.

\begin{figure}[t]
\centering
\includegraphics{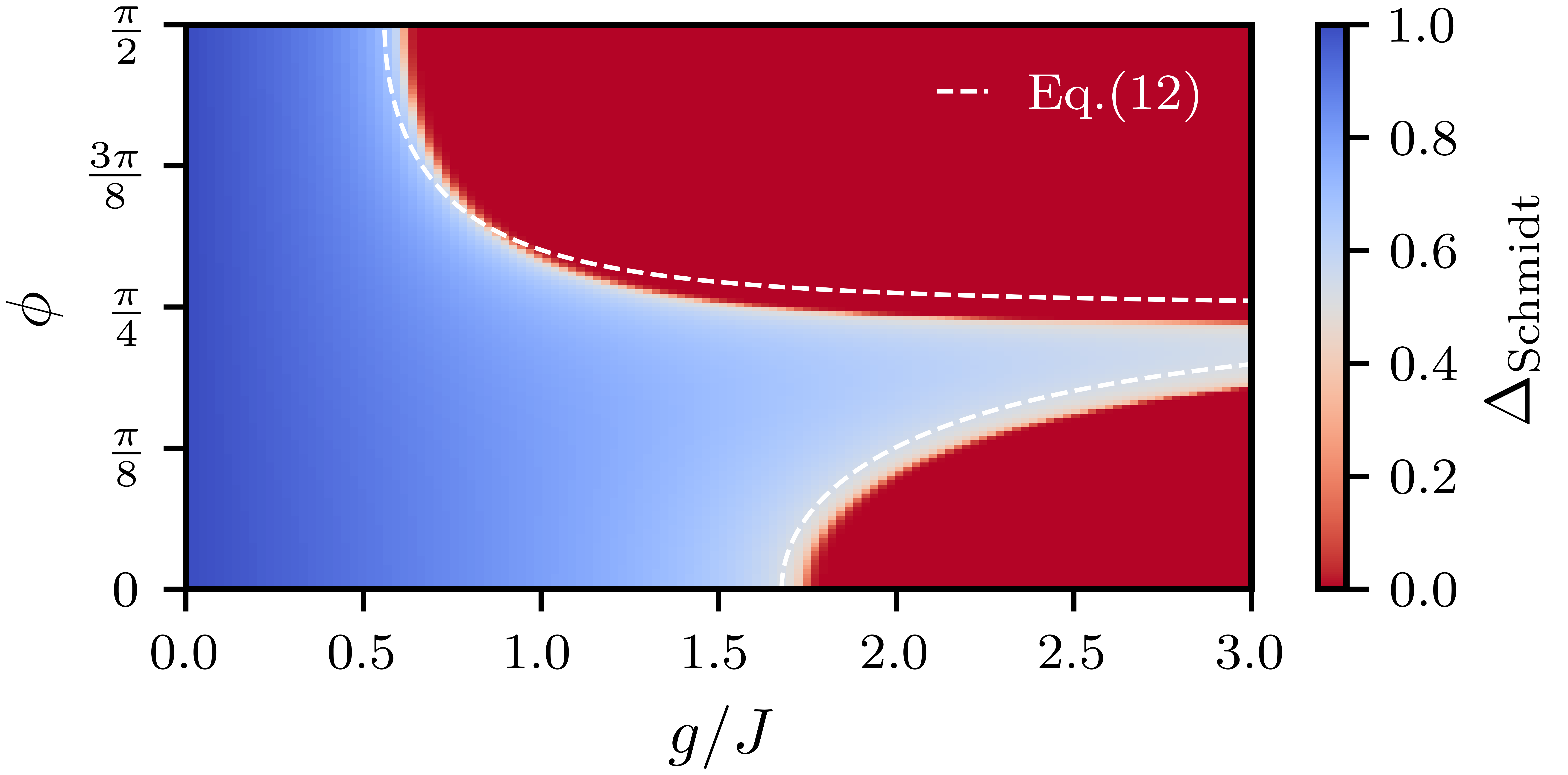}
\caption{(color online) Schmidt gap as a function of $g/J$ and $\phi$, calculated for a finite system with 256 spins. The dashed white lines represent predictions made with the free-fermion Hamiltonian (Eq.~\eqref{eq:Bulk_fermion_bogo_de_gennes}). Other parameters: $\Omega_{0}=\omega_{0}=2.5J$. \label{fig:schmidth_gap0}}
\end{figure}

We map out the criticality lines of the phase diagram separating different gapped phases by computing the values of $g$ where the single-particle gap of the Hamiltonian closes, 
\begin{equation}
g_{c} \left(\phi \right)=
\begin{cases}
\dfrac{1}{2}\sqrt{\dfrac{\omega_{0}\left(\Omega_{0}+2J\right)}{\cos\left(2\phi\right)},} & 0\leq\phi<\pi/4,\\
\\
\dfrac{1}{2}\sqrt{\dfrac{\omega_{0}\left(\Omega_{0}-2J\right)}{\left|\cos\left(2\phi\right)\right|},} & \pi/4<\phi\leq\pi/2.
\end{cases}\label{eq:critical_coupling_FF-1}
\end{equation}
These values are represented by dashed white lines in Fig.~\ref{fig:schmidth_gap0}. For these values of $g_c$, the system is in a gapless phase with the band touching at $k=0$ at zero energy for $\phi\in\left[0,\nicefrac{\pi}{4}\right)$ and at $k=\pi$ for $\phi\in\left(\nicefrac{\pi}{4},\nicefrac{\pi}{2}\right]$. However, we note that there is no transition for finite values of $g$ when $\phi=\pi/4$. Moreover, as $g$ is increased, we observe that the region that in-between the two critical lines, which is connected to the decoupled limit, is increasingly shorter. It is interesting to consider the $g\gg\Omega_0,\omega_0,J$ limit, where the full Hamiltonian is exactly solvable and the excitation spectrum of the single-particle Hamiltonian is given by,
\begin{equation}
E_k = \begin{cases}
\pm 2g\left(\cos\phi + \sin\phi\right),\\[1ex]
\pm 2g\left(\cos\phi - \sin\phi\right).
\end{cases}
\end{equation}
We observe that the excitations do not have any dispersion as the kinetic energy term is suppressed in this limit. Using this, we can compute the ground-state energy as a function of $\phi$,
\begin{equation}
E_{\rm GS}/N= \begin{cases}
- 4g\cos\phi,& 0\leq\phi\leq\pi/4,\\[1.5ex]
-4g\sin\phi,& \pi/4<\phi\leq\pi/2.
\end{cases}
\end{equation}
We observe that as \( g/J \rightarrow +\infty \), the two critical lines merge and give rise to a first-order quantum critical point at \( \phi = \pi/4 \), as the first derivative of the ground-state energy exhibits a discontinuity at this angle. We will return later to this point, when discussing the order parameters for $\phi>\pi/4$ and for $\phi<\pi/4$. This first order transition only occurs in this limit, and if we perturbatively consider the remaining terms present in the Hamiltonian, there is the opening of an energy gap, which separates the red region for $\phi>\pi/4$ from the other red region for $\phi<\pi/4$ in Fig.~\ref{fig:schmidth_gap0}.

It is important to emphasize that the analysis conducted so far does not rely on our particular choice of $\Omega_0$ and $\omega_0$, as long as $\Omega_0>2J$ and $\omega_0>0$. Changing these parameters only changes the value of the critical lines and does not introduce new gapped phases in the phase diagram.
\subsection{Entanglement Spectra}
Now, we obtain the phase diagram, but taking into consideration the quartic terms of Eq.~\eqref{eq:quartic_terms}. We map it through the entanglement properties of the ground-state wave function, which is obtained via DMRG calculations. Specifically, we analyze both the scaling of the entanglement entropy (a discussion deferred to appendix~\ref{appendix:phasediagram_interacting}) and the eigenvalues ($\lambda_\alpha$) of the reduced density matrix $\rho_\ell$, obtained by tracing out the degrees of freedom of half of the chain: $\rho_{\ell} = \text{Tr}_{\bar{\ell}} \left( \ket{\text{GS}}\bra{\text{GS}} \right)$. These eigenvalues, also known as the entanglement spectrum, have been shown to provide valuable information. As explored in several works~\cite{Li2008,Pollmann2010,tiwari2024quantumrestoredsymmetryprotected}, the degeneracies of the entanglement spectrum serve as reliable indicators of both topological order~\cite{Levin2006} and symmetry-protected topological phases~\cite{PhysRevB.96.165124}. For instance, the degeneracy pattern of the reduced density matrix eigenvalues can uniquely identify the eight topological phases of one-dimensional interacting fermionic systems that exhibit time-reversal symmetry and particle-number parity conservation~\cite{Turner2011, Fidkowski2011}.

Motivated by this, we use the entanglement spectrum to obtain the phase diagram of the full interacting model described by Eq.~\eqref{eq:fermionic_hamiltonians}. In the non-interacting limit, this model supports both a trivial phase and a topological phase hosting Majorana edge modes, as discussed in more detail in the appendix~\ref{appendix:majonara_edge_nI}. For an arbitrary value of $\phi$, the Hamiltonian in Eq.~\eqref{eq:fermionic_hamiltonians} conserves only the fermionic parity, as such, the system then exhibits two phases~\cite{PhysRevB.96.165124,Zeng2019}: a trivial phase and a non-trivial (topological) phase in which the entanglement spectrum is doubly degenerate. For interaction strengths below the critical value, the spectrum is generally non-degenerate. However, tuning $g$ or $\phi$ across the phase transition induces a change to a doubly degenerate spectrum.
\begin{figure}[t]
\begin{centering}
\includegraphics{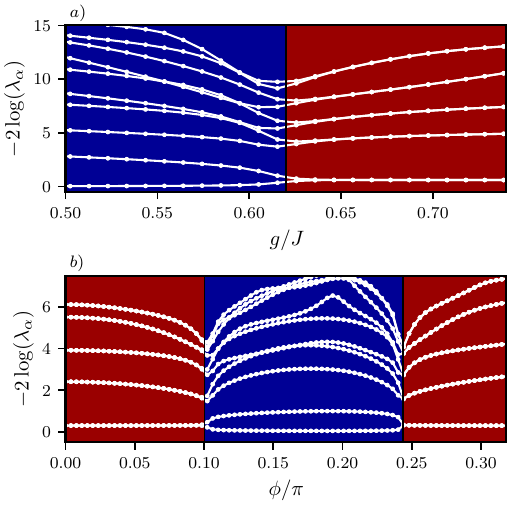}
\par\end{centering}
\caption{(color online) Entanglement spectrum as a function of the interaction strength $g$ for $\phi=\pi/2$ ($a)$) and as a function of the geometrical angle $\phi$ for $g=2.0J$ ($b)$). The blue region corresponds to the trivial phase. The red region is the topologically non-trivial phase. Other parameters: $\Omega_{0}=\omega_{0}=2.5J$. \label{fig:entanglement_spectrum}}
\end{figure}
This behavior is illustrated in Fig.~\ref{fig:entanglement_spectrum}, where we plot the logarithm of the entanglement spectrum as a function of the interaction strength for $\phi=\pi/2$ (panel $a)$) and as a function of $\phi$ for $g=2.0J$ (panel $b)$). The regime in which the entanglement spectrum exhibits double degeneracy corresponds to the topological phase, while the other regime is trivial. To determine the full phase diagram, we use the Schmidt gap, defined as
\begin{equation}
    \Delta_{\text{Schmidt}} = \left| \lambda_1 - \lambda_2 \right|,
\end{equation}
where $\lambda_1$ and $\lambda_2$ are the two largest eigenvalues of the reduced density matrix. Fig.~\ref{fig:schmidth_gap0} shows the Schmidt gap as a function of interaction strength and $\phi$, obtained from DMRG simulations. As in the non-interacting model, two distinct regions appear, with a phase boundary that depends on both the interaction strength and $\phi$. The resulting phase diagram closely resembles that obtained from the free-fermionic approximation, as evidenced by the similarity of the critical lines in both cases (see Fig.~\ref{fig:schmidth_gap0}). This suggests that the interaction term primarily renormalizes the system parameters.

\section{Role of $\phi$ in Shaping Magnetic Order}
\label{sec:magnetic_order_universality_class}

The free-fermionic approximation allowed us to derive an approximate analytical expression for the phase boundaries of the model. We found that the angle $\phi$ shifts the critical coupling $g_c$, and for certain specific values of $\phi$, no transition occurs for any finite $g$ and, remarkably, this behavior persists even when the quartic interaction term in Eq.~\ref{eq:fermionic_hamiltonians} is included.

The presence of two topological phases in the fermionic model, as identified by the entanglement degeneracies, guarantees that the corresponding spin model also exhibits two distinct phases~\cite{Zeng2019}: (i) a symmetric phase, in which the ground state preserves the spin $\mathbb{Z}_2$ symmetry (that is the ground-state is an eigenstate of the operator $\mathcal{P}$ defined in Eq.~\eqref{eq:symmetry_op}); and (ii) a phase in which the state spontaneously breaks the $\mathbb{Z}_2$ symmetry. These two phases correspond, respectively, to the trivial and topological phases of the fermionic model~\cite{Zeng2019}.

This is analogous to the physics of the Kitaev chain~\cite{Kitaev2001, Leijnse2012}, which is mapped to the TFIM via the JW transformation~\cite{Lieb1961, Pfeuty1970, Greiter2014}. In this mapping, the trivial phase of the Kitaev chain corresponds to the paramagnetic phase of the TFIM model, while the topological phase corresponds to a magnetically ordered phase arising from spontaneous breaking of the spin $\mathbb{Z}_2$ symmetry.

\subsection{Formation of in-plane Stripe order in the Broken Phase}

We begin by discussing and characterizing the magnetic order that develops in the $x-y$ plane in the broken phase, that is the red region in Fig.~\ref{fig:schmidth_gap0}.

\begin{figure}[t]
    \centering  \includegraphics{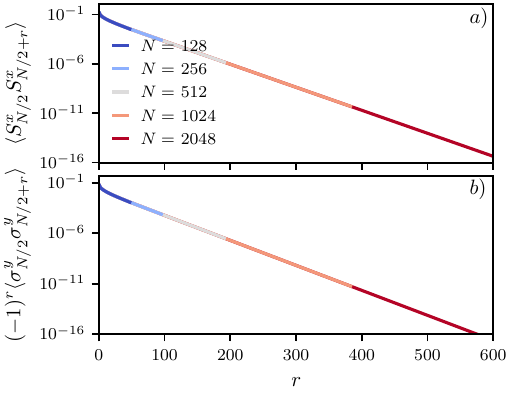}
    \caption{(color online) Spin-spin correlation function between the middle spin and a spin at distance $r$ for $g<g_c$. Panel $a)$ corresponds to $\phi=0$, while panel $b)$ corresponds to $\phi=\pi/2$. Other parameters: $\omega_0 = \Omega_0 = 2.5J$.}
    \label{fig:spin-spin_correlations1}
\end{figure}

In this regime, all excitations are gapped and consist of spin flips in chain A and spinless fermionic excitations in chain B, with a dispersion relation \( \varepsilon_k = \Omega_0 + 2J\cos k \), as obtained via the JW transformation. As the coupling $g$ increases from the decoupled limit, the ground state continues to exhibit long-range order along the $z$-direction, due to the dominance of the energy level splittings $\Omega_0, \omega_0 > g$. For values of $g$ and $\phi$ below the critical value, the ground-state of the system remains in a spin-\( \mathbb{Z}_2 \) symmetric phase. In this phase, operators that connect sectors with different eigenvalues of the parity operator \( \mathcal{P} \) have vanishing expectation values. As a result, neither chain A nor chain B exhibits in-plane magnetic order, \( \langle S_n^+ \rangle = \langle \sigma_n^+ \rangle = 0 \). Moreover, the in-plane spin–spin correlation functions decay exponentially, $\langle S^+_{n} S^-_{n+r}\rangle \sim \exp(-r/\xi)$ with $\xi$ denoting the correlation length, as illustrated in Fig.~\ref{fig:spin-spin_correlations1}. These correlations emerge at finite values of $g/J$, and their ferromagnetic or antiferromagnetic character is determined by the geometric phase $\phi$. For $0 < \phi < \pi/4$, ferromagnetic correlations develop along the $x$ direction in both chains A and B. In contrast, for $\pi/4 < \phi < \pi/2$, antiferromagnetic correlations arise: along the $x$ direction in chain B and along the $y$ direction in chain A.

As $g \to g_c$, the correlation length diverges in the thermodynamic limit following a power law,
\begin{equation}
    \xi \propto |g - g_c(\phi)|^{-\nu},
\end{equation}
with critical exponent $\nu = 1$. We obtain this in Appendix~\ref{appendix:critical_exponents}, where we also find that the characteristic time scale diverges with the same scaling, implying a dynamical critical exponent $z = 1$, so that at criticality the system is governed by a relativistic field theory~\cite{Sachdev2011}.
 
\begin{figure}[t]
    \centering
    \includegraphics{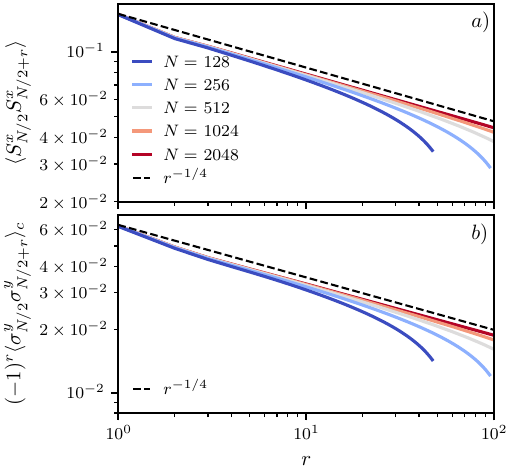}
    \caption{(color online) Spin-spin correlation between the middle spin and a spin at distance $r$ for $g=g_c$. Panel $a)$ correspond to the case $\phi = 0$, while  panel $b)$ corresponds to $\phi=\pi/2$. We observe that correlation function are decaying algebraically, which preludes the formation long-range order in the plane. Other parameters: $\omega_0 = \Omega_0 = 2.5J$.}
    \label{fig:spin-spin_correlations2}
\end{figure}

As the correlation length diverges, the in-plane spin-spin correlation functions decay following a power-law~\cite{M_E_Fisher_1968}
\begin{equation}
\lim_{r\rightarrow \infty}\langle S_{N/2}^{x}S_{N/2+r}^{x} \rangle \propto \frac{e^{-r/\xi}}{r^{(d+z-2+\eta)}},
\end{equation}
where $\eta$ is the anomalous critical exponent. We determine $\eta$ by fitting the spin-spin correlation function near the critical coupling as shown in panels $b)$ and $e)$ in Fig.~\ref{fig:spin-spin_correlations2}. Our results are consistent with $\eta = \nicefrac{1}{4}$. Although finite-size effects impose a cutoff by limiting the correlation length to scale as $N^\nu$, increasing the size of the system extends the range over which this power-law behavior is observed.

\begin{figure}[t]
    \centering
    \includegraphics[width=\columnwidth]{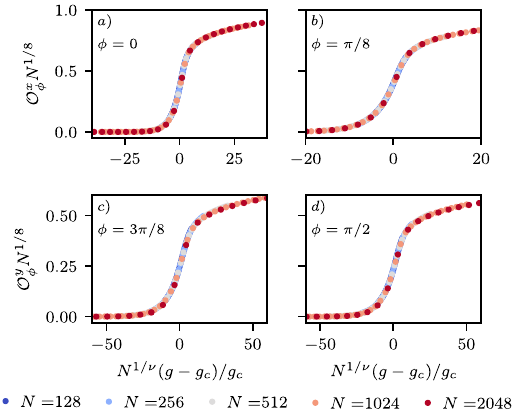}
    \caption{(color online) Finite-size scaling analysis of the order parameter, using the \emph{ansatz} $\mathcal{O}^{\alpha}_\phi = N^{-\beta/\nu} f ( \left(g-g_c \right) N^{1/\nu})$, plotted as a function of interaction strength for different values of $\phi$. A data collapse is achieved with the critical exponents $\beta = 1/8$ and $\nu = 1$ for all $\phi$. The corresponding critical interaction strengths are $g_c(\phi = 0) = 1.73(1)J$, $g_c(\phi = \pi/8) = 2.14(5)J$, $g_c(\phi = 3\pi/8) = 0.69(7)J$, and $g_c(\phi = \pi/2) = 0.60(5)J$. Other parameters are fixed at $\Omega_0 = \omega_0 = 2.5J$.
    \label{fig:order_parameter_fss}}
\end{figure}

We can define a suitable order parameter to distinguish between the symmetric and broken phase through the asymptotic value of the spin-spin correlation function,
\begin{equation}
\left(\mathcal{O}^\alpha \right)^2 = \lim_{r\rightarrow+\infty}\expected{S_{N/2}^{\alpha}S_{N/2+r}^{\alpha}},
\end{equation} 
with $\alpha=x/y$. This parameter is zero in the $\mathbb{Z}_2$ symmetric phase, but acquires a finite value when the ground-state spontaneously breaks the symmetry. 
We now fix a particular value of $\phi$ and analyze the behavior of the order parameter as the interaction strength $g$ is varied across the critical point for a fixed value of $\phi$. We note that one could also consider fixing a given strength of the coupling and tuning the angle instead to interpolate between the two different phases. We compute the order parameter from the square root of the spin-spin correlation function evaluated at a distance $r^\ast = 3N/4$. This choice ensures that we capture the long-distance behavior while minimizing boundary effects. Performing a finite-size scaling analysis as shown in Fig.~\ref{fig:order_parameter_fss}, we obtain that in the thermodynamic limit, the order parameter is expected to behave as
\begin{equation}
\mathcal{O}^\alpha \propto 
    \begin{cases}
    0, & g < g_c,\\
    (g-g_c)^{\beta}, & g>g_c,
    \end{cases}
\end{equation}
for all values of $\phi$, the critical exponent $\beta=1/8$. These critical exponents we have extracted correspond to the ones of the two-dimensional classical Ising model. Importantly, these exponents are independent of $\phi$, which is expected because the Hamiltonian retains its $\mathbb{Z}_2$ symmetry and the effective dimensionality $D = d + z = 2$ for all values of $\phi$. This confines the criticality of the system to the Ising universality class. Moreover, as we explore in the appendix~\ref{appendix:phasediagram_interacting}, the central charge observed from the entanglement entropy scaling at the critical point, together with the computed critical exponents, confirms that the quantum critical point is described by the Ising Conformal Field Theory.

\begin{figure}[t]
\centering
\includegraphics{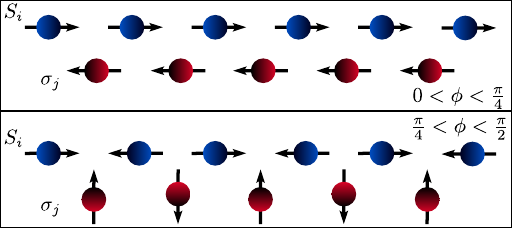}
\caption{(color online) Representation of the system's ground-state configuration in the regime $g\gg\Omega_{0},\omega_{0},J$. \label{fig:final_order}}
\end{figure}

As already observed from the spin–spin correlation functions, the magnetic order that emerges after the quantum phase transition strongly depends on the geometric angle $\phi$. Specifically, we identify two distinct in-plane antiferromagnetic stripe configurations, schematically illustrated in Fig.~\ref{fig:final_order}. For $0 \leq \phi < \pi/4$, stripes form along the $x$ direction, with the spins in both chains A and B also aligned along this axis; we refer to this state as Stripe-XX. In contrast, for $\pi/4 < \phi < \pi/2$, the stripes are again formed along the $x$ direction, but while the spins in chain B point along $x$, the spins in chain A are oriented along $y$. We refer to this state as Stripe-XY. These correspond to two distinct states of matter that break the Hamiltonian’s symmetries in different ways. Although both spontaneously break the spin $\mathbb{Z}_2$ symmetry-making them indistinguishable in the analysis based on the entanglement spectrum-they break the lattice translation symmetry differently: the Stripe-XX phase has a magnetic unit cell of the same size as the system’s unit cell, whereas the Stripe-XY phase has a magnetic unit cell that is the double.

In the limit $g\rightarrow+\infty$, the ground-state is solely determined by the Hamiltonian in Eq.~\eqref{eq:interaction_hamiltonian}, which confirms the type of order observed. Namely, the ground-state corresponds to the following product-state,
\begin{equation}
\ket{\text{GS} } =
\begin{cases}
\ket{\underbrace{\downarrow_{x}\cdots\downarrow_{x}}_{\text{A}};\underbrace{\uparrow_{x}\cdots\uparrow_{x}}_{\text{B}}},& 0<\phi<\pi/4,\\[1ex]
\ket{\underbrace{\uparrow_{y}\downarrow_{y}\cdots\uparrow_{y}\downarrow_{y}}_{\text{A}};\underbrace{\uparrow_{x}\downarrow_{x}\cdots\uparrow_{x}}_{\text{B}}},& \pi/4<\phi<\pi/2.
\end{cases}
\label{eq:gs_assymptotic_limit}
\end{equation}

As discussed in Sec.~\ref{sec:boundaries_phase_diagram}, in the limit $g/J \to +\infty$, a first-order quantum phase transition occurs at $\phi = \pi/4$. Now, this is also evidenced by a discontinuous change in the order parameter: varying $\phi$ through $\pi/4$ results in an abrupt switch between $\mathcal{O}^x$ and $\mathcal{O}^y$.

\subsection{Induced Finite vector spin Chirality}

The finite geometric angle in $\mathcal{H}_{\rm int}$ not only sets the spin orientation in the stripe phase but also favors non-collinear magnetic order. This is revealed through the emergence of a finite vector spin chirality within each chain, defined as
\begin{equation}
\boldsymbol{\kappa}_{n,n+1} = \boldsymbol{S}_n \times \boldsymbol{S}_{n+1}.
\end{equation}
In what follows, we focus on the expectation value of its \emph{z}-component, $\kappa^z_{n,n+1} = \hat{z} \cdot (\boldsymbol{S}n \times \boldsymbol{S}{n+1})$, since the $x$- and $y$-components vanish in the $\mathbb{Z}_2$-symmetric phase. In classical spin systems, this quantity captures the relative twisting of neighboring spins, providing a measure of the local sense of rotation.

\begin{figure}[t] 
\centering 
\includegraphics{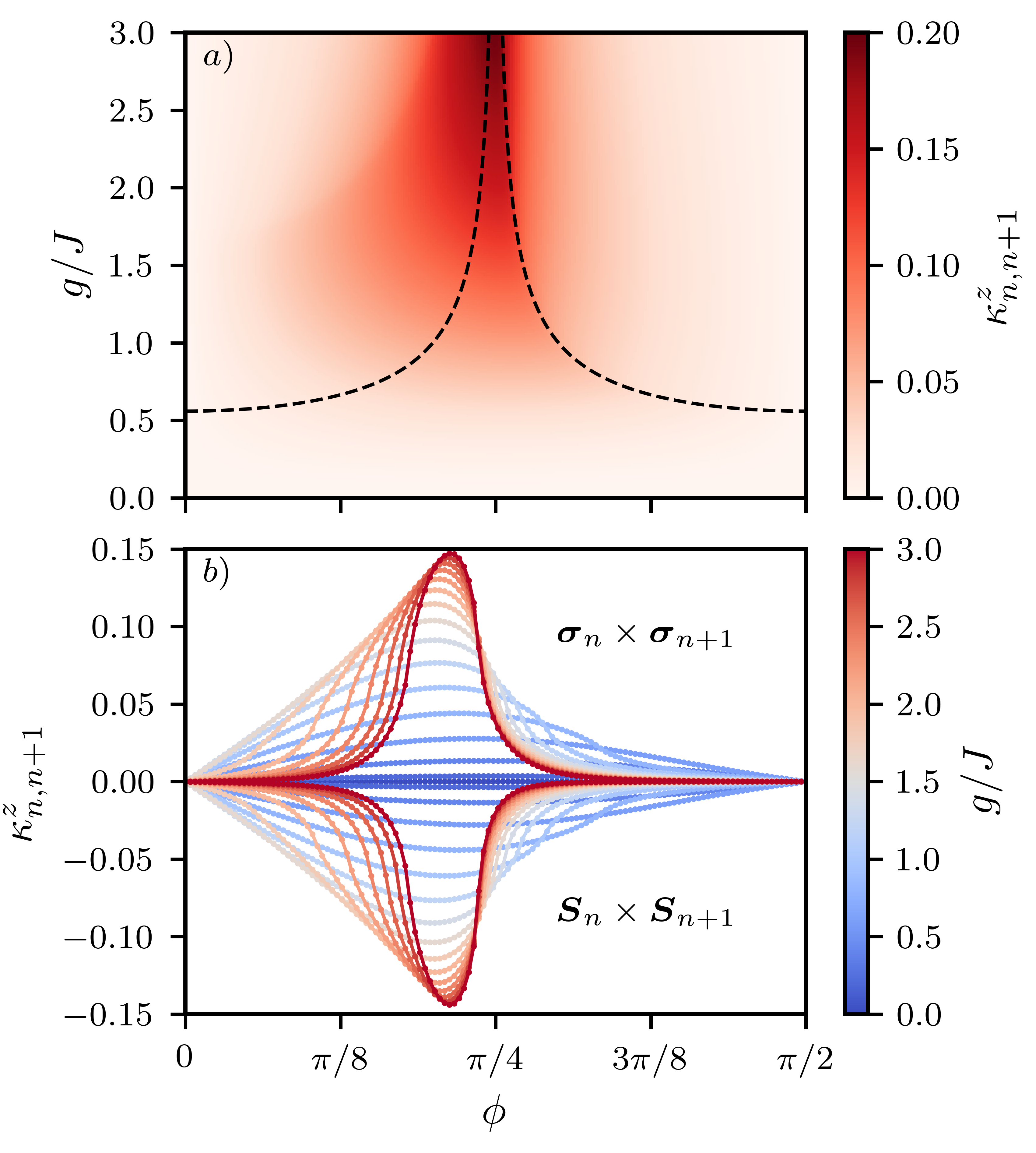} \caption{(Color online) $a)$ - Vector spin chirality along nearest-neighbor bonds in chain A, computed using the free-fermion model, as a function of the geometric interaction angle $\phi$ and interaction strength $g/J$. $b)$ - Vector spin chirality for both chains A and B as a function of $\phi$ for selected values of $g/J$. Other parameters: $\Omega_0 = \omega_0 = 2.5J$.
\label{fig:Vector_spin_chirality}} 
\end{figure}

Figure~\ref{fig:Vector_spin_chirality} illustrates the behavior of $\kappa^z$ as a function of the interaction strength $g$ and the geometric angle $\phi$. Panel $a)$ shows the full dependence of $\kappa^z$, obtained using the free-fermion approximation introduced in Sec.~\ref{sec:boundaries_phase_diagram}, while panel $b)$ displays its variation with $\phi$ for fixed values of $g/J$. The latter results were obtained via DMRG simulations on chains with $512$ spins, with the vector spin chirality averaged over all bulk bonds. 

Panels $a)$ and $b)$ show that a net vector spin chirality develops smoothly once the interaction becomes finite. This crossover occurs without criticality, as the interaction explicitly breaks inversion symmetry. The ground state thus evolves continuously from the fully polarized down state at $g=0$ to one with finite vector spin chirality, whose magnitude increases with $g$ and exhibits a nontrivial dependence on $\phi$. Notably, $\kappa^z$ vanishes at $\phi=0$ and $\phi=\pi/2$, reaching a maximum between these two limits. This behavior reflects the structure of the interaction Hamiltonian, which reduces to
\begin{equation}
\begin{aligned}
\mathcal{H}_{\text{int}} &= 4g \sum_{n} \sigma^x_n\left(S^x_{n} + S^x_{n+1}\right), && \phi = 0, \\
\mathcal{H}_{\text{int}} &= 4g \sum_{n} \sigma^y_n\left(S^x_{n+1} - S^x_{n}\right), && \phi = \pi/2,
\end{aligned}
\end{equation}
and thus does not favor non-collinear configurations within a chain, explaining the vanishing chirality. Panel $b)$ further shows that the chirality is symmetric on chains A and B, highlighting the distinct directional character induced by a finite $\phi$ on each chain. Reversing the angle, $\phi \to -\phi$, would simply invert the chirality on each chain. Within the symmetric phase, the chirality grows monotonically with $g$ as long as the $\mathbb{Z}_2$ symmetry is preserved. In the broken phase, however, collinear in-plane order is stabilized and the chirality tends to vanish. 

The free-fermion approximation captures the same qualitative behavior, with the vector spin chirality maximized near $\phi=\pi/4$. Under the Jordan–Wigner transformation, the vector spin chirality maps to fermionic operators:
\begin{equation}
\begin{aligned}
\hat{z} \cdot\left(\boldsymbol{\sigma}_n \times \boldsymbol{\sigma}_{n+1} \right) &\rightarrow \tfrac{i}{2} \left(c^\dagger_{n} c_{n+1} - c^\dagger_{n+1} c_{n} \right) e^{i\pi b^\dagger_{n+1} b_{n+1}}, \\
\hat{z} \cdot\left(\boldsymbol{S}_n \times \boldsymbol{S}_{n+1} \right) &\rightarrow \tfrac{i}{2} \left(b^\dagger_{n} b_{n+1} - b^\dagger_{n+1} b_{n} \right) e^{i\pi c^\dagger_n c_n}.
\end{aligned}
\end{equation}
From this, it follows that a nonzero chirality arises only when the band structure is asymmetric, $E_{k,\lambda} \neq E_{-k,\lambda}$. For $\phi=0$, the JW transform of the vector spin chirality is odd under time-reversal, while for $\phi=\pi/2$ it is also odd under the combined action of time-reversal and fermion parity on chain B. In both cases, $\kappa^z$ must vanish.

These findings show that the chiral interaction not only shifts the position of the quantum critical point but also qualitatively modifies the magnetic order in the symmetric phase. In the symmetric phase, for $\phi=0$, correlations primarily develop along the $x$-axis (preluding the formation of the stripe order). By contrast, a finite geometric angle $\phi$ induces non-collinear spin arrangements within each chain, manifested as a finite $\kappa^z$.

\begin{figure}[t]
    \centering
    \includegraphics{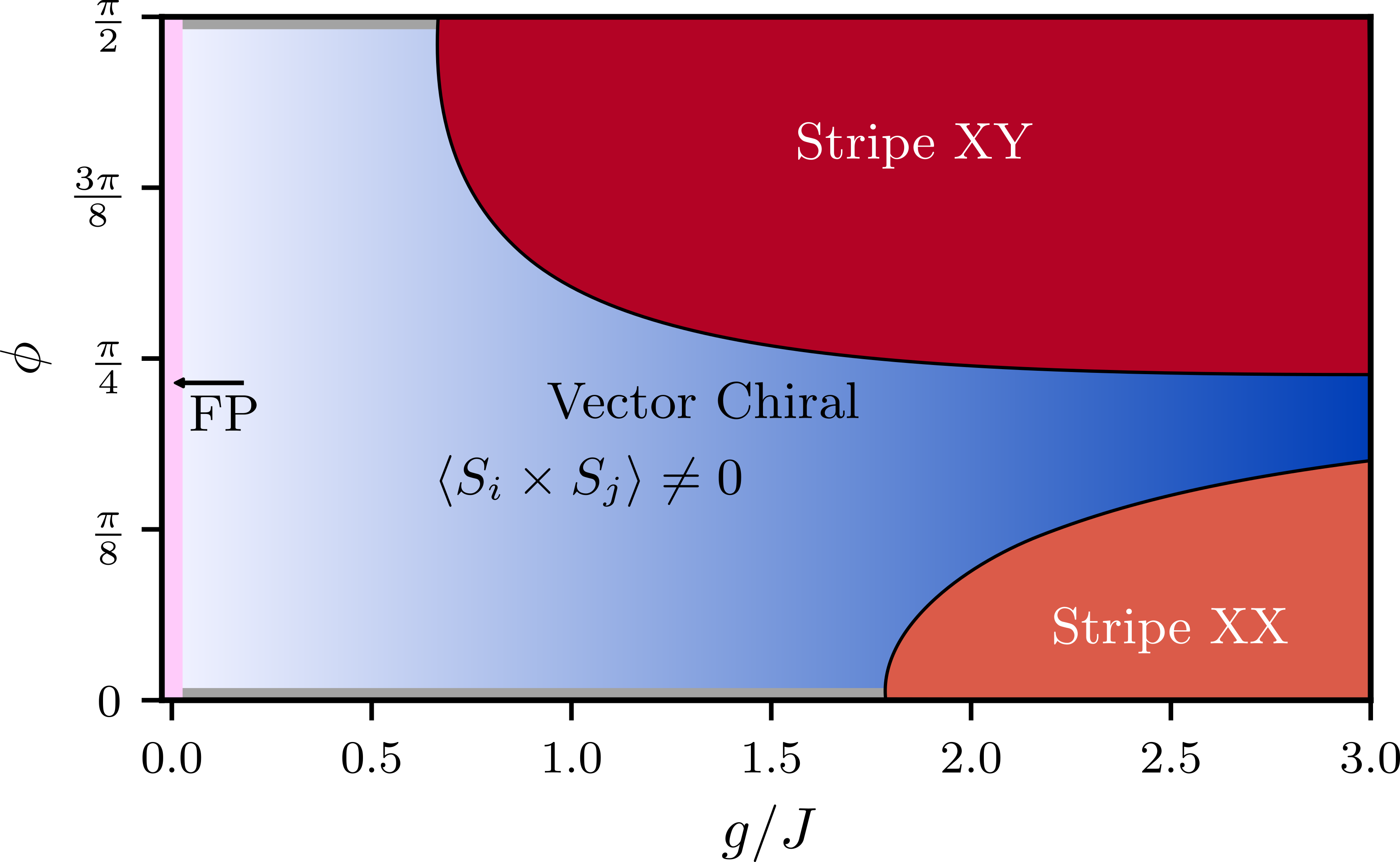}
\caption{(color online) Ground-state phase diagram as a function of $g/J$ and $\phi$. For $g = 0$, the ground state of the system is fully polarized in the down direction for all values of $\phi$. In the symmetric phase, at $\phi \ne 0$ and $\phi = \pi/2$, interactions induce a non-zero vector spin chirality with opposite orientation along chains A and B. In the broken-symmetry phase, two distinct stripe phases emerge, depending on the value of the geometric angle $\phi$.}
    \label{fig:summary_phase_diagram}
\end{figure}

\section{Conclusions}
\label{sec:conclusions}

In this work, we have established the ground-state phase diagram of two spin chains coupled by a chiral interaction tunable both in strength and, crucially, through the geometric phase angle $\phi$ between the chains. We uncover a rich phase diagram where the inter-chain angle serves as a powerful control parameter for stabilizing nontrivial magnetic textures, including finite vector spin chirality and distinct collinear orders, as summarized in Fig.~\ref{fig:summary_phase_diagram}.

First, we delineated the phase diagram by analyzing the entanglement spectrum. This reveals two distinct regions separated by a critical line: a symmetric phase with well-defined spin parity and a symmetry-broken phase where parity is spontaneously broken in the thermodynamic limit The transition between them is driven by the inter-chain coupling and depends nontrivially on both $g$ and $\phi$, as we have obtained analytically. Remarkably, the geometric phase $\phi$ not only shifts the critical boundary but can also fully suppress the transition when tuned appropriately.

We then characterized the phases through the expectation values of the spin–spin correlators in the $x-y$ plane. These confirm the absence of long-range in-plane magnetic order in the symmetric phase, while in the symmetry-broken phase reveal two regimes where in-plane stripe patterns emerge with different orientations. Remarkably, the stripe orientation is entirely controlled by the geometric phase $\phi$. We also extracted the critical exponents of the transition and found them to be consistent with the Ising universality class, independently of whether $g$ or $\phi$ is used to tune across the critical point.

In addition, we show that the interaction induces a finite vector spin chirality with opposite orientations in the two chains. This effect appears immediately when the interaction is turned on, without any critical behavior. The vector spin chirality vanishes at $\phi = 0$ and $\pi/2$, where the interaction instead favors collinear order. Within the symmetric phase, it grows with interaction strength but persists only over shrinking $\phi$ intervals that prevent crossing the phase transition. By contrast, in the symmetry-broken phase, the vector spin chirality is strongly suppressed because collinear in-plane order develops instead.

All these regimes are summarized in Fig.~\ref{fig:summary_phase_diagram}: (i) down-state polarization at zero inter-chain coupling, (ii) chain-specific vector spin chirality emerging beyond a critical coupling $g$, vanishing at $\phi = 0$ and $\pi/2$, and (iii) two types of antiferromagnetic stripes whose orientation depends on $\phi$. 

\begin{acknowledgments}
The DMRG calculations have been performed using the ITensor library~\cite{Fishman2022,Fishman2022a}. We thank João P. Santos Pires and Dhruv Tiwari for enlightening discussions and comments on the early versions of the draft. R.D.S acknowledges funding from Erasmus$+$ (Erasmus Mundus program of the European Union). The numerical calculations presented in this work were performed in the GRID FEUP High Performance Computing infrastructure.
\end{acknowledgments}

\bibliographystyle{apsrev4-1}
\bibliography{bib}

\appendix
\section{Entanglement Scaling of the ground-state}
\label{appendix:phasediagram_interacting}
In this appendix, we examine the entanglement scaling~\cite{Osborne2002,Vidal2003,RevModPhys.80.517,Laflorencie2016} of the ground-state wave function, which provides further information on the quantum criticality~\cite{Gu2003,Vidal2003,Verstraete2004} of the model.

\begin{figure}[b]
\begin{centering}
\includegraphics{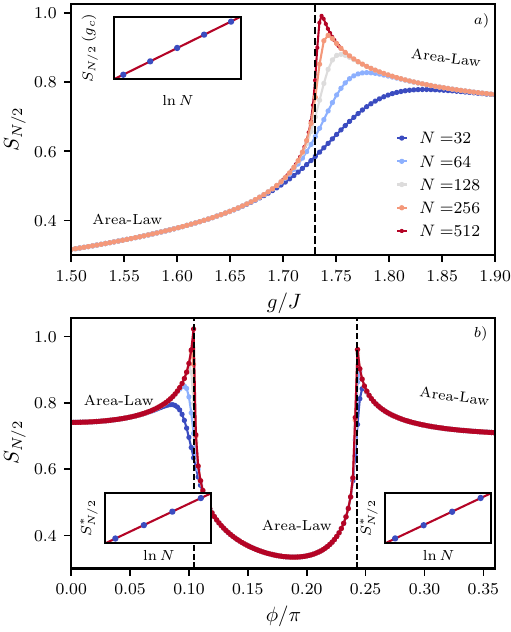}
\par\end{centering}
\caption{(Color online) Half-system entanglement entropy for different system lengths as a function of $g/J$ at $\phi=0$ [panel $a)$], and as a function of $\phi$ at $g=2.0J$ [panel $b)$]. As shown in the inset, the entanglement entropy scales logarithmically with the total system size at the critical points. The fitting reveals the central charge from the CFT description $c = \left(0.490 \pm 0.05 \right)$, which aligns with the Ising (and Majonara) CFT with $c= \nicefrac{1}{2}$. Away from the critical phase, it follows the area-law behavior. Other parameters: $\Omega_0 = \omega_0=2.5J$. \label{fig:EE_area_to_log_to_area}}
\end{figure}

First, we consider the von Neumann entropy (or entanglement entropy) of the reduced density matrix, $\rho_{\ell}$, associated to a continuous region of the system with linear size $\ell$,
\begin{equation}
    S_{\ell} = -\text{Tr}\left( \rho_{\ell} \ln \rho_{\ell} \right),
    \label{eq:EE_formula}
\end{equation}
where the reduced density matrix is obtained by tracing out the degrees of freedom in the complementary region, $\rho_{\ell} = \text{Tr}_{\bar{\ell}} \left( \ket{\text{GS}}\bra{\text{GS}} \right)$, where $\ket{\text{GS}}$ is the system's ground state. We take this region to be half of the chain and first examine how the entanglement entropy scales with total system size. In gapped ground states with short-range interactions and hoppings~\cite{Wolf2006, Hastings_2007}, the entanglement entropy follows an area law, scaling with the boundary of the subsystem: $S_\ell \propto \ell^{d-1}$, where $d$ is the spatial dimension. In contrast, critical systems usually show deviations from this behavior. In one-dimensional critical systems described by a (1 + 1)d conformal field theory (CFT), the entanglement entropy scales instead logarithmically with $\ell$, as proved by Cardy and Calabrese~\cite{Calabrese2009}. Fig.~\ref{fig:EE_area_to_log_to_area} presents the entanglement entropy for different system sizes as a function of the interaction strength. The transition in the scaling of the entanglement entropy from an area law to logarithmic form allows us to identify the critical value $g_c$ marked by the black dashed line in the figure. We observe that the interactions renormalize this value compared to the prediction from Eq.~\eqref{eq:critical_coupling_FF-1}. We can determine the central charge of the (1+1)d CFT that describes the critical point by fitting the entanglement entropy at $g_c$ as a function of the total size of the system to the Cardy and Calabrese formula~\cite{Calabrese2009}
\begin{equation}
\label{eqn:scft}
S_{N/2}=\dfrac{c}{6}\ln\left(\dfrac{4N}{\pi}\right)+s_0,
\end{equation}
where $s_0$ is a non-universal constant that does not scale with the total system size. The fitting is shown in the inset of Fig.~\ref{fig:EE_area_to_log_to_area}, and from this we extract a central charge of $c = \left(0.490 \pm 0.05\right)$ which matches the central charge of the Ising CFT (and Majorana\footnote{The central charge of the CFT remains unchanged under the JW transformation, hence the Ising and Majorana CFTs share the same central charge as they are interrelated through this mapping.}) CFT. These together with the extracted critical exponents corroborate that our model belongs to the 2D Ising model universality class, and its critical point is therefore described by the Ising CFT.

\section{Critical exponents}
\label{appendix:critical_exponents}

\begin{figure}
    \centering
    \includegraphics{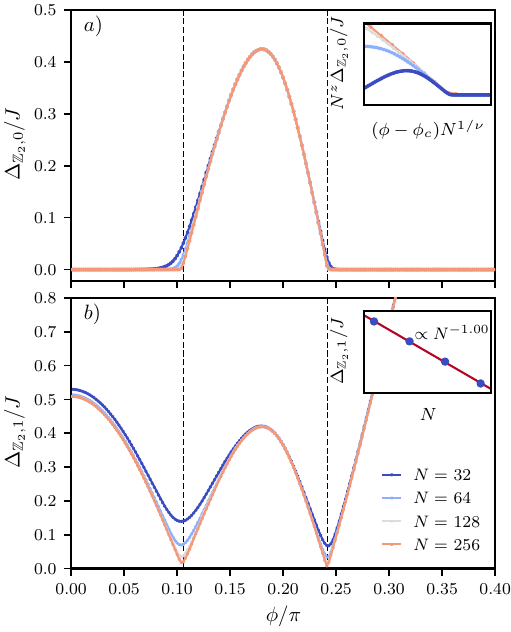}
    \caption{(color online) $a)$ – Dependence of $\Delta_{\mathbb{Z}_2,0}$ on the interaction angle $\phi$. The inset shows a finite-size scaling analysis near the second critical point. The data collapse was obtained using $\phi_c/\pi \simeq 0.2498$ and $\nu = 1$. $b)$ – Dependence of $\Delta_{\mathbb{Z}_2,1}$ on the interaction angle $\phi$. In the inset, we show the dependence of the gap on the system size at the critical point, which allows extraction of the dynamical critical exponent $z = (1.00 \pm 0.01)$. Other parameters: $\Omega_0 = \omega_0 = 2.5J$ and $g = 2.0J$.}
    \label{fig:gap_exponent}
\end{figure}

Here, we extract the critical exponents by studying the behavior of the energy gap near the critical point and the second derivative of the ground-state energy. We start by looking to the energy gap between the ground states in each parity sector. This is done by imposing quantum number conservation in the DMRG algorithm~\cite{Fishman2022}. We determine the energy difference between the ground states in the two distinct symmetry sectors of the spin $\mathbb{Z}_2$ symmetry:
\begin{equation}
\Delta_{\mathbb{Z}_2,0} = E_{\text{GS},\mathcal{P}=-1} - E_{\text{GS},\mathcal{P}=+1},
\label{eq:gap_z2_1}
\end{equation}  
where $E_{\text{GS},\mathcal{P}=\pm 1}$ denotes the ground-state energy in the even ($\mathcal{P}=+1$) and odd ($\mathcal{P}=-1$) sectors. We also consider the energy difference between the ground-state energy in the even sector and the first excited state in the odd sector:
\begin{equation}
\Delta_{\mathbb{Z}_2,1} = E_{\text{1}^{\text{st}},\mathcal{P}=-1} - E_{\text{GS},\mathcal{P}=+1}.
\label{eq:gap_z2_2}
\end{equation}  
In Fig.~\ref{fig:gap_exponent}, we plot these quantities as we tune the angle $\phi$ for an interaction strength of \(g=2.0J\). Panel $a)$ shows the difference in ground-state energies between the two symmetry sectors, while panel $b)$ displays the energy difference between the ground state in the even sector and the first excited state in the odd sector. For $g=2.0J$, we observe the existence of two critical values of $\phi$, highlighted by the dashed vertical lines. In panel $a)$, we see that for $\phi$ between the two critical values, the ground-state energy in the even sector is lower than in the odd sector. This indicates that the ground state is unique and lies in the even sector. As the angle is tuned to either of the critical values, we observe that $\Delta_{\mathbb{Z}_2,0}$ tends to zero. In this regime, in the thermodynamic limit, the ground states in both parity sectors become degenerate, and thus the system's ground state is no longer necessarily an eigenstate of the $\mathcal{P}$ operator — indicating spontaneous breaking of the spin $\mathbb{Z}_2$ symmetry. We observe that away from the critical points, the system remains gapped, consistent with the area-law scaling of the entanglement entropy discussed in appendix~\ref{appendix:phasediagram_interacting}.

\begin{figure}
    \centering
    \includegraphics{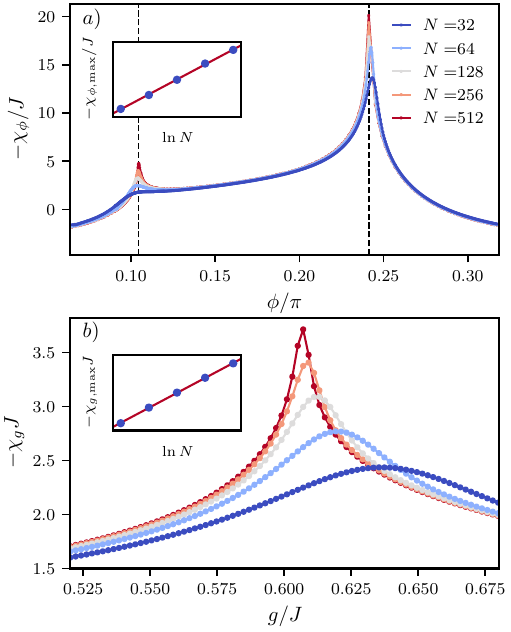}
    \caption{(color online) Second derivative of the ground-state energy density as a function of the interaction angle $a)$ and the interaction strength $b)$ for different system sizes. Panel $a)$ corresponds to the case $g = 2.0J$, while panel $b)$ corresponds to $\phi = \pi/2$. The inset shows the dependence of the maximum of the second derivative on system size. The fit was performed using Eq.~\ref{eq:xi_g_ffs_fit}. Other parameters: $\Omega_0 = \omega_0 = 2.5J$.}
    \label{fig:gs_second_dev}
\end{figure}

We study the scaling of the energy gap at the critical points by computing it using different system sizes and perform a finite-size scaling analysis to extract the universal critical exponent governing the vanishing of the characteristic energy scale near the critical point~\cite{Sachdev2011}. We expect that in the thermodynamic limit,
\begin{equation}
\Delta_{\mathbb{Z}_2} \propto \left| \phi - \phi_c \right|^{z\nu}, 
\label{eq:gap_power_law}
\end{equation}
where $\nu$ characterizes the divergence of the correlation length $\xi$ near the critical point and $z$ is the dynamical critical exponent. In non-relativistic quantum models, the absence of Lorentz invariance prevents spatial and temporal dimensions from being treated on equal footing. Consequently, the characteristic length and time scales diverge differently at criticality, and $z$ accounts for this difference. For critical points described by a relativistic field theory, one expects $z = 1$. For a finite system, the correlation length cannot diverge at criticality, being simply proportional to the system size. As such, the energy gap at the critical point should scale with system size as
\begin{equation}
\Delta_{\mathbb{Z}_2} (\phi_c, N) \propto N^{-z}.
\end{equation}
More generally, for $\phi$ near the critical value, we expect the gap to obey the following finite-size scaling \emph{ansatz}:
\begin{equation}
\Delta_{\mathbb{Z}_2} (\phi, N) = N^{-z} f\left((\phi - \phi_c) N^{1/\nu} \right),
\end{equation}
where $f$ is a universal scaling function.

In the insets of Fig.~\ref{fig:gap_exponent}, we show both $\Delta_{\mathbb{Z}_2,0}$ and $\Delta_{\mathbb{Z}_2,1}$ near the critical point. In the inset of panel $a)$, we demonstrate that it is possible to collapse the data for different system sizes using the scaling \emph{ansatz}. This collapse was only achieved by setting $z = 1$, which required assuming that the correlation length exponent $\nu$ is one at the estimated critical point. In the inset of panel $b)$, we highlight that the energy gap closes linearly with system size, corroborating that the critical exponent $z$ is consistent with unity. Thus, our finite-size scaling analysis supports the following critical exponents:
\begin{equation}
z = 1, \quad \nu = 1.
\end{equation}

We can further substantiate the universality class by extracting the critical exponent $\alpha$ from the structure of the ground-state energy density, $\varepsilon_{\text{GS}}$, which at the critical point exhibits a non-analytic behavior. In particular, the second derivatives of the ground-state energy density with respect to the tuning parameters
\begin{equation}
\chi_g = \frac{1}{N}\frac{\partial^2 \varepsilon_{\text{GS}}}{\partial g^2}, \quad
\chi_\phi = \frac{1}{N}\frac{\partial^2 \varepsilon_{\text{GS}}}{\partial \phi^2},
\end{equation}
diverge at the critical point. Finite-size scaling analysis reveals that at the critical point, this divergence depends logarithmically on the system size:
\begin{equation}
\begin{aligned}
\left. \chi_g \right|_{g=g_c} &= q \ln N + q_0, \\
\left. \chi_\phi \right|_{\phi=\phi_c} &= q \ln N + q_0,
\end{aligned}
\label{eq:xi_g_ffs_fit}
\end{equation}
where $q$ and $q_0$ are non-universal constants. This is shown in the insets of both panels of Fig.~\ref{fig:gs_second_dev}. In the thermodynamic limit, this result implies the following logarithmic divergences:
\begin{equation}
\chi_g \propto \ln \left| g - g_c \right|, \quad
\chi_\phi \propto \ln \left| \phi - \phi_c \right|.
\end{equation}
This behavior is analogous to that found in the TFIM~\cite{Fradkin2021}, where the tunable parameter is an external magnetic field. In the TFIM, one can construct a quantum–classical mapping that relates the logarithmic divergence in \(\chi_g\) to that of the specific heat at the critical temperature of the classical two-dimensional Ising model on a square lattice. In both cases, the absence of a dominant power-law divergence in $\chi_g$ and $\chi_\phi$ indicates that the critical exponent $\alpha$ is zero. This result, along with the previous ones, supports that the quantum phase transition in our system belongs to the Ising universality class.

\section{Breakdown of Linear Spin-Wave Theory\label{sec:SPIN-WAVE-THEORY}}
We explore and discuss the validity of a linear spin-wave theory (LSWT) treatment of the model near the ground-state for $g=0$, which corresponds to the fully-polarized down. This approach should be valid when the coupling strength, $g$, between the two spin chains is small compared to the other energy scales of the system.

In the framework of spin-wave theory~\cite{Holstein1940}, the spin ladder operators are mapped to bosonic ones following
\begin{equation}
\begin{aligned}
\sigma_{n}^{+}&=a_{n}^{\dagger}\sqrt{2S-a_{n}^{\dagger}a_{n}}, \quad
\sigma_{n}^{-}=\left(\sigma^+_n \right)^\dagger,\\
S_{n}^{+}&=b_{n}^{\dagger}\sqrt{2S-b_{n}^{\dagger}b_{n}}, \quad
S_{n}^{-}=\left( S^+_n \right)^\dagger,
\end{aligned}
\end{equation}
where $a^\dagger_n$ $( b^\dagger_n)$ and $a_n$ $(b_n)$ are bosonic operators that create/destroy a boson in the nth site of chain A(B), and $S$ corresponds to the total spin quantum number (in our case $S=\nicefrac{1}{2}$).
Assuming that the average excitation number per site is negligible, $\forall_{n},\left\langle S_{n}^{+}S_{n}^{-}\right\rangle \ll1$, an $1/S$ expansion can be carried out as follows, 
\begin{equation}
\begin{aligned}
\sigma_{n}^{+} & \simeq\sqrt{2S}\left(a_{n}^{\dagger}-\dfrac{1}{4S}a_{n}^{\dagger}a_{n}^{\dagger}a_{n}\right)+\mathcal{O}\left(S^{-3/2}\right),\\
S_{n}^{+} & \simeq\sqrt{2S}\left(b_{n}^{\dagger}-\dfrac{1}{4S}b_{n}^{\dagger}b_{n}^{\dagger}b_{n}\right)+\mathcal{O}\left(S^{-3/2}\right).
\end{aligned}
\end{equation}
Taking the first order of development and performing a Fourier transform of the bosonic operators, we obtain a quadratic bosonic Hamiltonian with non-null bosonic pairing terms,
\begin{equation}
\begin{aligned}
\mathcal{H}_{\text{A}} & =2S\omega_{0}\sum_{k}a_{k}^{\dagger}a_{k},\\
\mathcal{H}_{\text{B}} & =2S\sum_{k}\Omega_{k}b_{k}^{\dagger}b_{k},\\
\mathcal{H}_{\text{int}} & =2S\sum_{k}\left(g_{k}a_{-k}^{\dagger}\left(b_{k}^{\dagger}+b_{-k}\right)+\text{h.c.}\right),
\end{aligned}
\label{eq:bosonic_hamiltonian_HP}
\end{equation}
where $g_{k}=2ge^{ika/2}\cos\left(\phi+ k/2\right)$. This Hamiltonian can be diagonalized by a Hopfield-Bogoliubov transformation~\cite{Cortese2017,Colpa1978},
\begin{equation}
A_{k}=u_{k}a_{k}+v_{k}a_{-k}^{\dagger}+w_{k}b_{k}+z_{k}b_{-k}^{\dagger}.
\end{equation}
where $u_{k},v_{k},w_{k}$ and $z_{k}$ represent the Hopfield coefficients
satisfying the normalization condition $\left|u_{k}\right|^{2}-\left|v_{k}\right|^{2}+\left|w_{k}\right|^{2}-\left|z_{k}\right|^{2}=1$,
such that $\left[A_{k},A_{q}^{\dagger}\right]=\delta_{k,q}$. The diagonalization of the Hamiltonian is reduced to the following eigenvalue problem,
\begin{equation}
\boldsymbol{\mathcal{L}}_{k}\boldsymbol{U}_{k}=\Omega_{k}\boldsymbol{U}_{k},
\end{equation}
where $\boldsymbol{U}_{k}=\begin{pmatrix}u_{k} & v_{k} & w_{k} & z_{k}\end{pmatrix}$,
and,
\begin{figure}[t]
\centering
\includegraphics{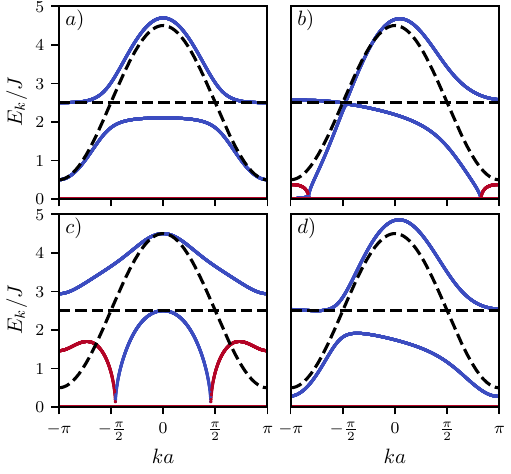}
\caption{(color online) Hybridization of the modes for different parameters of the system. The dashed black curves correspond to the bare dispersion of the spin waves. The solid lines correspond to the dispersion relation of the hybridized modes. The blue (red) lines corresponds to the real (imaginary) component. In $a)$ $\left(g,\phi\right)=\left(0.4J,0\right)$,
in $b)$ $\left(g,\phi\right)=\left(0.5J,\pi/4\right)$, in $c)$
$\left(g,\phi\right)=\left(J,\pi/2\right)$ and in $d)$ $\left(g,\phi\right)=\left(0.6J,\pi/8\right)$.
Other parameters: $\Omega_{0}=2.5J$ and $\omega_{0}=2.5J$.\label{fig:(color-online)-Hybridisation_bosonic}}
\end{figure}
\begin{equation}
\boldsymbol{\mathcal{L}}_{k}=2S\left(\begin{array}{cccc}
\Omega_{k} & g_{-k}^{\ast} & 0 & g_{k}\\
g_{-k} & \omega_{0} & g_{-k} & 0\\
0 & -g_{-k}^{\ast} & -\Omega_{k} & -g_{k}\\
-g_{k}^{\ast} & 0 & -g_{k}^{\ast} & -\omega_{0}
\end{array}\right).
\end{equation}
Observe that the matrix above is not Hermitian. However, this is a well-known property of quadratic bosonic Hamiltonians that do not conserve particle number~\cite{Wang2019}. In the specific case $\phi=0$, it is possible to obtain in closed form the eigenvalues of the Hopfield-Bogoliubov matrices,
\begin{equation}
\begin{cases}
E_{k}^{(1)}=\sqrt{\dfrac{1}{2}\left(\Omega_{k}^{2}+\omega_{0}^{2}-\Upsilon_{k}\right)},\\
\\
E_{k}^{(2)}=\sqrt{\dfrac{1}{2}\left(\Omega_{k}^{2}+\omega_{0}^{2}+\Upsilon_{k}\right)},
\end{cases}
\end{equation}
where $\Upsilon_{k}=\sqrt{\left(\Omega_{k}^{2}-\omega_{0}^{2}\right)^{2}+16\omega_{0}\Omega_{k}\left|g_{k}\right|^{2}}$ and $E^{(\nu)}$ correspond to the dispersion of band $\nu$. For an arbitrary $\phi$, only a numerical result is possible. In Fig.~\ref{fig:(color-online)-Hybridisation_bosonic}, the dispersion relation is depicted for different parameters, as seen in the lower panel, imaginary modes start to appear for some values of the coupling parameter. When such is the case, the system is said to be dynamically unstable~\cite{McDonald2018}, which is expected for the modes satisfying the condition,
\begin{figure}[t]
\centering
\includegraphics{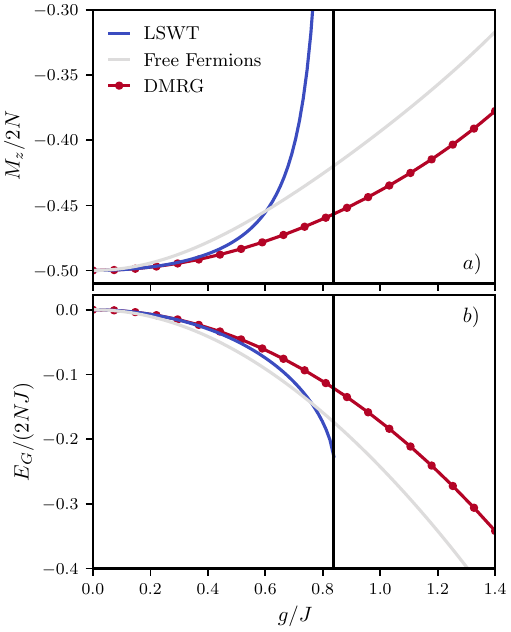}
\caption{(color online) Comparison of results from the LSWT shown in blue, with those from the free-fermionic approximation in grey, and DMRG depicted in red. The black vertical line corresponds to the $g^\text{LSWT}_c$ predicted by Eq.~\eqref{eq:instability_lswt} from where the LSWT is no longer applicable. $a)$ - Dependence of the total magnetization of the system per spin with the interaction strength $g$ is depicted. $b)$ - Ground-state energy per spin as a function of the interaction strength $g$. The DMRG calculations were obtained with $512$ spins. Other parameters: $\Omega_{0}=2.5J$, $\omega_{0}=2.5J$ and $\phi=0$.\label{fig:DMRG_vs_HP}}
\end{figure}
\begin{equation}
\left|g_{-k}\right|^{2}+\left|g_{k}\right|^{2}=\dfrac{\omega_{0}\Omega_{k}}{2}.
\label{eq:instability_lswt}
\end{equation}
In that case, we can determine the validity of the Hamiltonian by computing the onset of instability. If $k_{c}$ is the first unstable mode, we can determine $g_{c}$ as a function of the parameters,
\begin{equation}
g^{\text{LSWT}}_{c} \left(\phi \right)=\sqrt{\dfrac{\omega_{0}\Omega_{k}}{8\left(1+\cos\left(2\phi\right)\cos(k_{c})\right)}}.\label{eq:Instability_bosonic}
\end{equation}
Similarly to the fermionic case, the counterrotating terms also give rise to pairing terms in this bosonic language. Nevertheless, there is a crucial difference between the Hamiltonian of Eq.~\eqref{eq:bosonic_hamiltonian_HP} and of Eq.~\eqref{eq:fermionic_hamiltonians}. First, the bosonic Hamiltonian must be non-negative to ensure stability. Furthermore, to diagonalize both Hamiltonians, the transformation must preserve the canonical (anti-)commutation relations. In the fermionic case, this is achieved by a unitary transformation, while in the bosonic case, they are para-unitary~\cite{Colpa1978} as the spectrum is obtained by diagonalizing an effective non-Hermitian Hopfield-Bogoliubov matrix ~\cite{Flynn2020,Chaudhary2021}. Thus, the transformation must satisfy~\cite{McClarty2022}
\begin{equation}
\begin{aligned}\boldsymbol{U}_{k}^{\dagger}\boldsymbol{\mathcal{L}}_{k}\boldsymbol{U}_{k} & =\boldsymbol{E}_{k},\\
\boldsymbol{U}_{k}\boldsymbol{\eta}_{k}\boldsymbol{U}_{k} & =\boldsymbol{\eta}_{k},
\end{aligned}
\end{equation}
where $\boldsymbol{E}_{k}$ is a diagonal matrix and $\boldsymbol{\eta}_{k}=\text{diag}\left(1,1,-1,-1\right)$. So \emph{a priori}, there is no guarantee that the spectrum is real, and in principle, for a dynamically unstable system~\cite{Wang2019,McDonald2018}, the spectrum may be complex for some value of the system parameters. Furthermore, the Bogoliubov-Hopfield transformation may not exist~\cite{Wang2019} for every pairing bosonic Hamiltonian, which is in contrast to the fermionic case. This is the case for the Hamiltonian in Eq.~\eqref{eq:bosonic_hamiltonian_HP} as $g\sim J$. The dynamical instability signals the invalidity of the first-order spin wave theory for considerable values of $g$. The reason is twofold: there is a phase transition that radically changes the nature of the ground state, so an analysis based on the spin-wave excitations around the fully polarized down-state state $\ket{\Psi_{0}}=\ket{\downarrow\cdots \downarrow}$ is no longer sufficient or valid; the low-excitation assumption breaks down, and so this could explain why the critical value for $g_{c}$ in Eq.~\eqref{eq:Instability_bosonic} is half of what is predicted with the free-fermionic theory, Eq.~\eqref{eq:critical_coupling_FF-1}. Not surprisingly, the band dispersion represented in Fig.~\ref{fig:(color-online)-Hybridisation_bosonic} matches that obtained with the JW transformation for $g=0$. In this regime, the energy dispersion can also be obtained by studying the one-body sector of the Hamiltonian. This corresponds to a product state between a one-particle excitation in either of the spin chains. For chain A, this corresponds to spin-flip at a site $n$:
\begin{equation}
\ket{1_{n,\text{A}},0_\text{B}}=\sigma_{n}^{+}\ket{\downarrow\cdots\downarrow},
\end{equation}
This single particle excitation as energy $\varepsilon = \omega_0$ with respect to the ground-state. For the chain B, the single-magnon excitations have a non-zero momenta $k$~\footnote{Their existence is guaranteed by translational invariance (under PBC) and the fact that, at  $g=0$ all eigenstates have a well-defined number of excitations, as $\left[\mathcal{H},\sum_{n}\sigma_{n}^{z}+S_{i}^{n}\right]=0$.},
 \begin{equation}
\begin{aligned}
\ket{0_{A},k_{n,B}}=\dfrac{1}{\sqrt{N}}\sum_{n,m=0}^{N-1}e^{ik_{1}n}S_{n}^{+}\ket{\downarrow\cdots\downarrow},
\end{aligned}
\end{equation}
with $k_n=\frac{2\pi}{N}n$, $n\in\mathbb{Z}_{N}$. These single-magnon states have a momentum-dependent energy: $\varepsilon_{k}=2J\cos(k)+\Omega_{0}$.

\section{Majonara Fermions in the non-interacting limit}
\label{appendix:majonara_edge_nI}

\begin{figure}[t]
\centering
\includegraphics{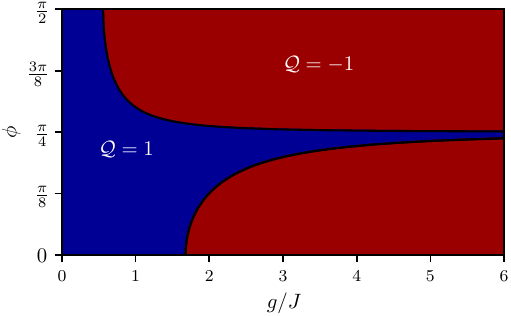}
\caption{(color online) Topological phase diagram of Hamiltonian of Eq.~\eqref{eq:Bulk_fermion_bogo_de_gennes}. The red (blue) region corresponds to the topological (trivial) phase. Other parameters: $\Omega_0 = \omega_0 = 2.5J$.\label{fig:phase_diagram_free_fermions}}
\end{figure}

In this appendix, we demonstrate that in the non-interacting limit the fermionic Hamiltonian considered in Eq.~\eqref{eq:fermionic_hamiltonians} has zero-edge Majorana quasiparticles. We begin by expressing the fermionic operators using Majorana operators:
\begin{equation}
\begin{aligned}
    \gamma_{1,A,n} &= c^\dagger_{n} +  c_{n}, \; \gamma_{2,A,n} = i\left( c_{n} - c^\dagger_{n}\right),\\
    \gamma_{1,B,n} &= b^\dagger_{n} +  b_{n}, \; \gamma_{2,B,n} = i\left( b_{n} - b^\dagger_{n}\right),\\
\end{aligned}
\end{equation}
where the Majonara operators satisfy the anti-commutation relations $\{ \gamma_{s,\alpha,n}, \gamma_{p,\beta,m} \} = 2\delta_{sp} \delta_{\alpha \beta}\delta_{nm}$. The non-interacting Hamiltonian in terms of these operators is given as
\begin{equation}
\begin{aligned}\mathcal{H}_{\text{A}} & =\dfrac{i \omega_{0}}{2}\sum_{n=0}^{N-1}\gamma_{1,A,n}\gamma_{2,A,n},\\
\mathcal{H}_{\text{B}} & =\dfrac{i}{2}\sum_{n=0}^{N-1}\left[\Omega_{0}\gamma_{1,B,n}\gamma_{2,B,n}-J\gamma_{2,B,n}\gamma_{1,B,n+1})\right],\\
\mathcal{H}_{\text{int}} & = i g\sum_{n=0}^{N-2} \gamma_{1,A,n} \left[ \gamma_{1,B,n+1} \sin\left(\phi \right) - \gamma_{2,B,n} \cos\left(\phi \right) \right]\\
& +\gamma_{2,A,n} \left[ \gamma_{1,B,n+1} \cos\left(\phi \right) - \gamma_{2,B,n} \sin\left(\phi \right) \right].
\end{aligned}
\end{equation}
Observe that in the limit $g\gg \omega_0,\Omega_0,J$, there are always two Majorana operators unpaired at the edge of the chain. The first Majorana corresponds to $\gamma_{1,B,0}$ as it does not appear in the Hamiltonian. The second mode varies with $\phi$. For example, when $\phi=0$, the Majonara $\gamma_{1,A,N-1}$ is unpaired, while for $\phi = \pi/2$, it is the Majorana $\gamma_{2,A,N-1}$. 
\begin{figure}[t]
\centering
\includegraphics{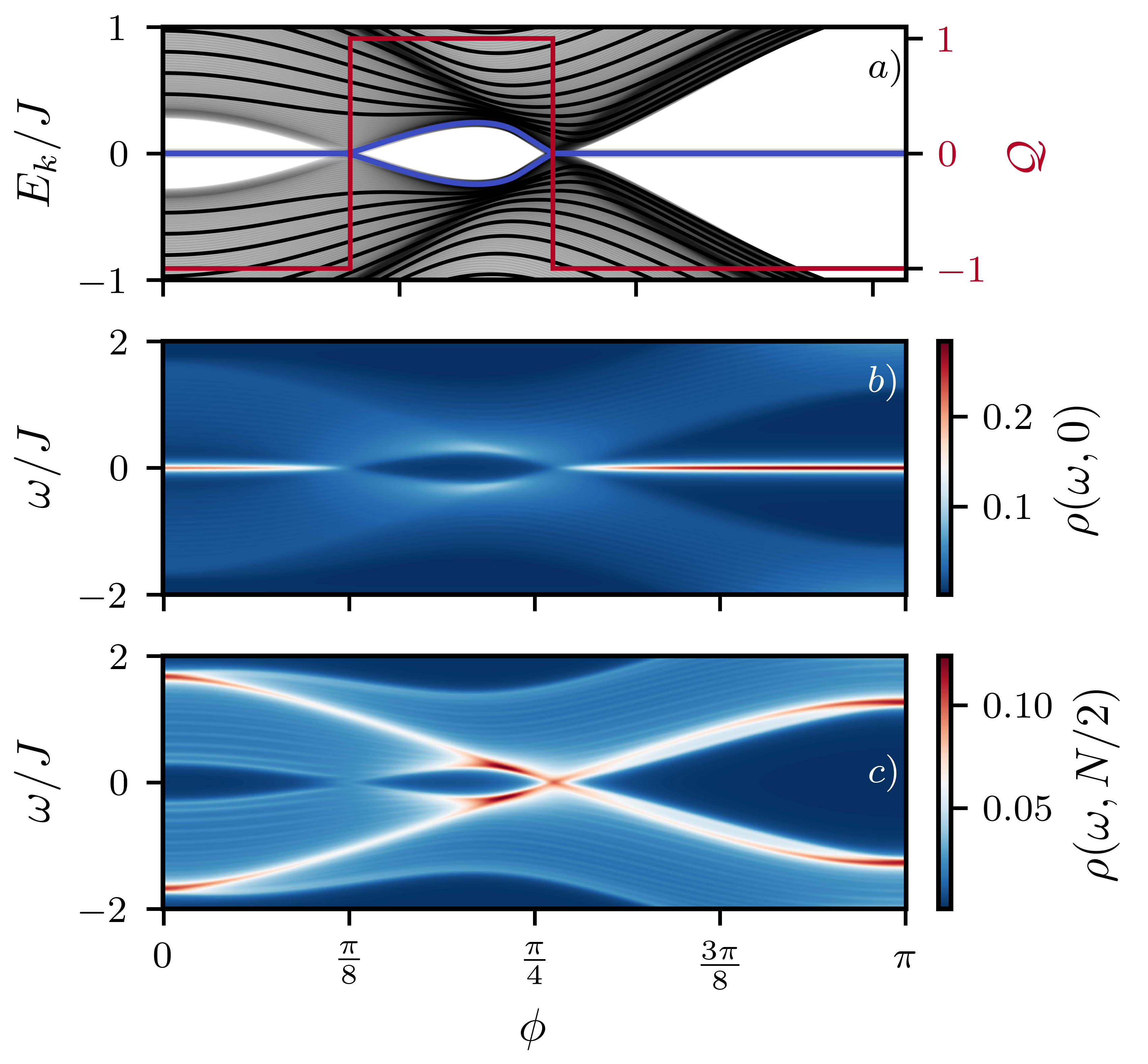}
\caption{(color online) $a)$ Energy spectrum as a function of interaction angle ($\phi$) for a system with OBC. The second vertical axis (red) shows the topological invariant $\mathcal{Q}$ as a function of $\phi$. The blue lines highlight two states that evolve into zero-energy modes when the bulk gap closes and the topological invariant changes sign. Panels $b)$ and $c)$: Local density of states at energy $\omega$ as a function of $\phi$, evaluated at the left edge $b)$ and the center $c)$ of the chain. Other parameters: $\Omega_0 = \omega_0 = 2.5J$ and $g = 2J$.
\label{fig:majonara_zero_mode}}
\end{figure}

As we have discussed in the main text, for a generic value of $\phi$, the single-particle Hamiltonian only satisfies the PHC of Eq.~\eqref{eq:phc}, so it belongs to class D in the periodic table of topological insulators and superconductors~\cite{PhysRevB.78.195125,Kitaev_periodic}; thus, the topological phases of the system are characterized by a $\mathbb{Z}_{2}$ topological invariant~\cite{PhysRevB.55.1142,RevModPhys.88.035005}. We consider the topological invariant introduced by Kitaev~\cite{Kitaev_periodic,RevModPhys.88.035005,Sato2017},
\begin{equation}
\mathcal{Q}=\text{sgn}\left(\text{Pf}\left[i\tilde{H}_{k=0}\right]\text{Pf}\left[i\tilde{H}_{k=\pi}\right]\right),\label{eq:Topological_Z2}
\end{equation}
where $\text{sgn}\left[\cdot\right]$ is the sign function, $\text{Pf}\left[\cdot\right]$ is the Pfaffian and $\tilde{H}_{k=0,\pi}$ is the single-particle Hamiltonian matrix in the Majorana basis evaluated at $k=0,\pi$, namely:
\begin{equation}
\tilde{H}_{k} =  
\begin{pmatrix}
\mathcal{I}_{2\times 2} & \mathcal{I}_{2\times 2}\\
i\mathcal{I}_{2\times 2} & -i\mathcal{I}_{2\times 2} 
\end{pmatrix}
H_{k}
\begin{pmatrix}
\mathcal{I}_{2\times 2} & -i\mathcal{I}_{2\times 2}\\
\mathcal{I}_{2\times 2} & i\mathcal{I}_{2\times 2} 
\end{pmatrix},
\end{equation}
where $\mathcal{I}_{2\times 2}$ is the two by two identity matrix. A value of $\mathcal{Q}=1$ corresponds to a trivial phase, while $\mathcal{Q}=-1$ indicates a topological phase, characterized by the presence of Majorana edge modes. Using this topological invariant, we obtain the topological phase diagram shown in Fig.~\ref{fig:phase_diagram_free_fermions}, whose phase boundaries are given by Eq.~\eqref{eq:critical_coupling_FF-1}. In panel $a)$ of Fig.~\ref{fig:majonara_zero_mode}, we show the single-particle energy spectrum as a function of $\phi$ for $g=2.0J$ in a system with OBC. The bulk gap closes and reopens, giving rise to a transition from a trivial phase (with $\mathcal{Q}=1$) to a topological phase (with $\mathcal{Q}=-1$) that supports Majorana zero modes at the chain edges~\cite{Kitaev2001,Beenakker2013}. These zero-energy states are localized at the edges of the system, which can be seen by computing the local density of states at energy $\omega$ and site $n$, defined as~\cite{PhysRevB.98.224512,PhysRevB.103.224505}
\begin{equation}  
    \rho(\omega, n) = -\dfrac{1}{\pi} \text{Im} \left( \text{Tr} \left[\omega + 0^+ - \mathcal{H} \right]^{-1} \right)_{n,n},  
\end{equation}  
where $\text{Im}(\cdot)$ denotes the imaginary part, $\text{Tr}(\cdot)$ represents the trace over the Nambu space and the sublattice degrees of freedom, and $\mathcal{H}$ is the single-particle Hamiltonian with OBC.

In panels $b)$ and $c)$ of Fig.~\ref{fig:majonara_zero_mode}, we show the local density of states at the left edge and at the center of the chain, respectively (the right edge exhibits similar behavior). The zero-energy states contribute significantly at the edges but are absent in the center, confirming their localized nature. In contrast, eigenstates with finite energy appear in both regions, indicating their delocalized character. These plots also reveal the hybridization of the trivial flat bands present for $g = 0$, where the localized states $c^\dagger_n \ket{\text{vac}}$ are eigenstates of the Hamiltonian.

The fermionic model studied in this manuscript differs from those previously examined in the literature. Essentially, our system constitutes a single Kitaev chain with an additional sublattice structure, distinguishing it from the so-called Kitaev ladder~\cite{Wu2012,Maiellaro2018,PhysRevResearch.2.013175}. Without the inter-chain coupling term, $g$, the two chains would decouple into two trivial band insulator models, lacking the necessary features to support Majorana fermions.
\end{document}